\documentclass[]{aa}

\usepackage{natbib}
\bibpunct{(}{)}{;}{a}{}{,}
\usepackage{graphicx}
\usepackage{txfonts}
\usepackage{xcolor}
\usepackage{epstopdf}
\usepackage{natbib}
\usepackage{scrextend}
\usepackage{multirow}
\usepackage{soul}

\def\msol{\hbox{\kern 0.20em $M_\odot$}}
\def\lsol{\hbox{\kern 0.20em $L_\odot$}}
\def\rsol{\hbox{\kern 0.20em $R_\odot$}}
\def\sr{\hbox{\kern 0.20em sr}}
\def\srmu{\hbox{\kern 0.20em sr$^{-1}$}}

\def\g{\hbox{\kern 0.20em g}}
\def\gmu{\hbox{\kern 0.20em g$^{-1}$}}
\def\kg{\hbox{\kern 0.20em kg}}
\def\pc{\hbox{\kern 0.20em pc}}

\def\mum{\hbox{\kern 0.20em $\mu$m}}
\def\mumd{\hbox{\kern 0.20em $\mu$m$^{-2}$}}
\def\cm{\hbox{\kern 0.20em cm}}
\def\m{\hbox{\kern 0.20em m}}
\def\km{\hbox{\kern 0.20em km}}
\def\nm{\hbox{\kern 0.20em nm}}

\def\s{\hbox{\kern 0.20em s}}
\def\h{\hbox{\kern 0.20em h}}
\def\sec{\hbox{\kern 0.20em sec}}
\def\min{\hbox {\kern 0.20em min}}
\def\smu{\hbox{\kern 0.20em s$^{-1}$}}
\def\smd{\hbox{\kern 0.20em s$^{-2}$}}
\def\an{\hbox{\kern 0.20em an}}
\def\anmu{\hbox{\kern 0.20em an$^{-1}$}}
\def\deg{\hbox{\kern 0.20em $^{\rm o}$}}
\def\yr{\hbox{\kern 0.20em yr}}
\def\yrmu{\hbox{\kern 0.20em yr$^{-1}$}}
\def\Myr{\hbox{\kern 0.20em Myr}}
\def\Mymu{\hbox{\kern 0.20em Myr$^{-1}$}}
\def\K{\hbox{\kern 0.20em K}}
\def\pcmu{\hbox{\kern 0.20em pc$^{-1}$}}
\def\pcmd{\hbox{\kern 0.20em pc$^{-2}$}}
\def\pcmt{\hbox{\kern 0.20em pc$^{-3}$}}
\def\kms{\hbox{\kern 0.20em km\kern 0.20em s$^{-1}$}}
\def\kmpd{\hbox{\kern 0.20em km$^{2}$}}
\def\kpc{\hbox{\kern 0.20em kpc}}
\def\cms{\hbox{\kern 0.20em cm\kern 0.20em s$^{-1}$}}
\def\erg{\hbox{\kern 0.20em erg}}
\def\ergs{\hbox{\kern 0.20em erg}}
\def\cmpd{\hbox{\kern 0.20em cm$^2$}}
\def\cmmd{\hbox{\kern 0.20em cm$^{-2}$}}
\def\cmms{\hbox{\kern 0.20em cm$^{-6}$}}
\def\cmpt{\hbox{\kern 0.20em cm$^3$}}
\def\cmmt{\hbox{\kern 0.20em cm$^{-3}$}}
\def\mpd{\hbox{\kern 0.20em m$^2$}}
\def\mmd{\hbox{\kern 0.20em m$^{-2}$}}
\def\mpt{\hbox{\kern 0.20em m$^3$}}
\def\mmt{\hbox{\kern 0.20em m$^{-3}$}}
\def\mujy{\hbox{\kern 0.20em $\mu$Jy}}
\def\mjy{\hbox{\kern 0.20em mJy}}
\def\Mj{\hbox{\kern 0.20em MJy}}
\def\jy{\hbox{\kern 0.20em Jy}}
\def\ghz{\hbox{\kern 0.20em GHz}}
\def\srmd{\hbox{\kern 0.20em sr$^{-1}$}}

\def \mum{$\mu$m}
\def\G{\hbox{\kern 0.20em G}}

\def\htwo{\hbox{H${}_2$}}
\def\h13cop{\hbox{H$^{13}$CO$^{+}$}}

\def\h2o{\hbox{H$_2$O}}

\begin{document}
\title{First Hot Corino detected around an Isolated Intermediate-Mass Protostar: Cep\,E-mm.}
%   \subtitle{I. A study of Cep\,E-mm}
\author{J.~Ospina Zamudio
          \inst{1}
          \and
          B.~Lefloch\inst{1}
          \and
          C.~Ceccarelli\inst{1}
          \and
          C.~Kahane\inst{1}
          \and
          C.~Favre\inst{2}
          \and
          A.~L\'opez-Sepulcre\inst{1,3}
      	  \and
          M.~Montarges\inst{4}}

   \institute{Univ. Grenoble Alpes, CNRS, IPAG, 38000 Grenoble, France\\
              \email{juan-david.ospina-zamudio@univ-grenoble-alpes.fr}
         \and
         	 INAF, Osservatorio Astrofisico di Arcetri, Largo E. Fermi 5, 50125 Firenze, Italy
         \and
             IRAM, 300 rue de la Piscine, 38406 Saint Martin d'Heres, France
         \and
         	 Institute of Astronomy, KU Leuven, Celestijnenlaan 200D B2401, 3001 Leuven, Belgium
         	 }
   \date{}
\abstract{
Intermediate-mass (I-M) protostars provide a bridge between the low- and high-mass protostars. Despite their relevance, little is known about their chemical diversity. }
{We want to investigate the molecular richness towards the envelope of I-M protostars and to compare their properties with other sources.}
{We have selected the isolated I-M Class 0 protostar Cep\,E-mm to carry out an unbiased molecular survey with the IRAM~30m telescope between 72 and 350~GHz with an angular resolution lying in the range 7\arcsec\ -- 34\arcsec. Our goal is to obtain a census of the chemical content of the protostellar envelope. These data were complemented with NOEMA observations  of the spectral bands 85.9 -- 89.6~GHz and 216.8 -- 220.4~GHz at an angular resolution of 2.3\arcsec\  and 1.4\arcsec\, respectively.}
{
30m spectra show bright emission of 13 O- and N-bearing complex organic molecules (COMs).
We identify three components in the spectral signature: an extremely broad line (eBL) component associated with the outflowing gas, a narrow line (NL) component associated with the cold envelope,  and a broad line (BL) component which traces the signature of a hot corino.
The NOEMA observations reveal Cep\,E-mm  as a binary protostellar system Cep\,E-A and Cep\,E-B separated by $\approx 1.7$\arcsec. Cep\,E-A dominates the  core continuum emission and powers the high-velocity jet associated with HH377.
Our interferometric maps show that COMs arises from a region of
$\approx 0.7\arcsec$ size around Cep\,E-A, and corresponds to the BL component detected with the IRAM 30m. We have determined the rotational temperature ($T_{rot}$) and the molecular gas column densities. Rotational temperatures were found to lie in the range $20\K$ -- $40\K$ with column densities ranging from a few $10^{15}\cmmd$ for O-species bearing, down to a few $10^{14}\cmmd$ for N-bearing species. Molecular abundances are similar to those measured towards other low- and intermediate-mass protostars. Ketene (H$_2$CCO) appears as an exception, as it is found significantly more abundant towards Cep\,E-A. High-mass hot cores are significantly less abundant in methanol and N-bearing species are more abundant by 2--3 orders of magnitude.}
{
Cep\,E-mm reveals itself as a binary protostellar system with a strong chemical differentiation between both cores. Only the brightest component of the binary is  associated with a hot corino. Its properties are similar to those of low-mass  hot corinos.
}

% 5 {} token are mandatory

%  \abstract

\maketitle
%
%-------------------------------------------------------------------

\section{Introduction}

The chemical composition of protostellar envelopes and their properties along the evolutionary stage of protostars is an important topic in Astrochemistry. Since the pioneering work by Cazaux et al. (2003) and Sakai et al. (2008), systematic chemical studies of solar-type protostars (see Ceccarelli et al. 2007, Caselli \& Ceccarelli 2012 for a review; also Lefloch et al. 2018) have identified two classes of objects. The first class corresponds to the so-called "hot corinos", i.e. sources which display a rich content in Complex Organic Molecules (COMs) in the central inner regions of the protostellar envelope (see Ceccarelli et al. 2007 for a review; also Taquet et al. 2015). Only a few hot corinos have been identified so far either with single dish or interferometric observations: IRAS16293-2422 (Cazaux et al. 2003; Bottinelli et al. 2004b; Jorgensen et al. 2011, 2016), IRAS2, IRAS4B (Bottinelli et al. 2007), IRAS4A (Bottinelli et al. 2004a, Taquet et al. 2015), HH212 (Codella et al. 2016), L483 (Oya et al. 2017), B335 (Imai et al. 2016), SVS13A (Bianchi et al. 2017), Serpens SMM1 and SMM4 (Öberg et al. 2011). We note that very few sources were investigated in a systematic manner so that the COM budget in hot corino sources is very inhomogeneous, making difficult a general picture to emerge.
 Hot corinos share some similarities with the hot cores observed around high-mass stars but they are not scaled-down versions of these. Bottinelli et al. (2007) showed that the abundances  of O-bearing species scaled to methanol are  higher  than those measured in hot cores by one to two orders of magnitude or more. The second chemical class of protostars corresponds to the so-called Warm Carbon Chain Chemistry (WCCC) sources, which have a rich content in C-chains but are poor in COMs.  A recent survey of a sample of 36 Class 0/I protostars of the Perseus molecular cloud complex by Higuchi et al. (2018) shows that the majority of the sources observed  have intermediate characters between these two distinct chemistry types.

In comparison, very little is known on Intermediate-mass (IM) protostars.  The first systematic study was carried out
by Crimier et al. (2010) and Alonso-Albi et al. (2010), who both investigated the physical and chemical properties of a sample of five Class 0 IM protostars (CB3-mm, Cep E-mm, IC1396 N BIMA 2, NGC 7129~FIRS 2, and Serpens FIRS 1). Crimier et al. (2010) derived the dust and gas temperature and density profiles of the protostellar envelopes from modelling their Spectral Energy Distribution  with the 1D radiative transfer code DUSTY. Comparing the physical parameters of the envelopes (density profile, size, mass) with those of the envelopes of low- and high-mass protostars, led them to concluded that the transition between the three groups appears smooth, and that the formation processes and triggers do not substantially differ. Alonso-Albi et al. (2010) studied the CO depletion and the N$_2$H$^+$ deuteration in the same source sample (and additionally towards L1641~S3~MMS1 and OMC2-FIR4). They found hints of CO underabundance by a factor of 2 with respect to the canonical value in the inner protostellar regions; they pointed out that high-angular resolution observations are needed to conclude about the origin of such a deficit, and the possible role of  outflows and the UV radiation from the star. The chemical properties of the source sample (molecular species and abundances) remain to be fully characterized. A few individual sources have been studied in detail (Fuente et al. 2007; Neri et al. 2007; Hogerheijde et al. 1999; Schreyer et al. 2002). In particular,  Fuente et al. (2005) reported the presence of a hot core in NGC 7129~FIRS 2.

Intermediate-mass Class 0 protostellar clusters are important too as they also provide the transition between the low-density groups of TTauri stars and the high-density clusters around massive stars. OMC2-FIR4 is one of the (if not the) best-studied prototype of this class. OMC-2 FIR4 is itself a young proto-cluster that harbours several embedded
low- and intermediate-mass protostars (Shimajiri et al. 2008; López-Sepulcre et al. 2013). Analysis of the CH$_3$OH emission lines observed with {\em Herschel} Space Observatory (Pilbratt et al. 2010) in the framework of the CHESS Large Program\footnote{http://chess.obs.ujf-grenoble.fr/} (Ceccarelli et al. 2010) led Kama et al. (2010) to suggest the presence of a hot core with kinetic temperatures above $100\K$ and around 400 au in size towards OMC-2FIR4. This is supported by the similarity of the molecular line spectrum of OMC-2 FIR4 and the hot corino source NGC 1333-IRAS4A, as observed with ASAI\footnote{http://www.oan.es/asai/} (Lefloch et al. 2018).  Observations with the Northern Extended Millimeter Array (NOEMA) at $5\arcsec$ resolution by Lopez-Sepulcre et al. (2013) have revealed a rather complex structure with the presence of several components of one or several solar mass each, with chemical differentiation and an ionized HII region just next to the OMC-2 dust envelope.
Hence, the chemical  properties of IM protostars remains to be characterized. The chemical diversity and its similarities
and differences with low- and high-mass sources remains to be established.

As part of an observational effort to address these questions, we have conducted a detailed molecular line survey of the isolated IM Class 0 protostar Cep\,E-mm (IRAS23011+6126) with the IRAM~30m telescope, complemented with observations with the IRAM interferometer NOEMA. Cep\,E-mm is
located in the Cepheus\,E molecular cloud at a nearby distance of $d$= $730\pc$ (Sargent 1977). Its luminosity is $\sim 100\lsol$ and its envelope mass is $35\msol$ (Lefloch et al. 1996; Crimier et al. 2010). Since its early discovery by Wouterloot \& Walmsley (1986) and Palla et al. (1993), subsequent studies confirmed the source to be an isolated intermediate-mass protostar in the Class 0 stage (Lefloch et al. 1996; Moro-Martin et al. 2001). The source drives a very luminous molecular outflow and jet (Lefloch et al. 2011, 2015), terminated by the bright Herbig-Haro object HH377 (Ayala et al. 2000).

The article is organized as follows. In Section 2, we present the observations carried out with the IRAM~30m telescope and the NOEMA interferometer. In Section 3, we present the results on the source multiplicity obtained from the  thermal dust emission, as mapped with NOEMA, and the status of the dust components. In Section 4, we present the results of a systematic search for complex organic molecules (COMs), chemically related species with the IRAM instruments, which shows the presence of a hot corino associated with one of the dust components. In Section 5, we discuss the physical structure of Cep\,E-mm, the evidence for chemical differentiation in the core, and we compare our results with those obtained towards a few typical objects, from low- to high-mass. Our conclusions are summarized in Section 6.

%--------------------------------------------------------------------
\section{Observations}

\subsection{Single-dish survey}

Observations on Cep\,E-mm were obtained with the IRAM 30\-m telescope near Pico Veleta (Spain) on coordinates $\alpha(2000)=\mathrm{23^h03^m12^s.8},\,\delta(2000)=\mathrm{61^{\circ}42'26''}$. Unbiased spectral coverage
was carried out at 3\,mm (80-116\,GHz), 2\,mm (129-173\,GHz), 1.3\,mm (200-300\,GHz) and 0.9\,mm (330-350\,GHz) in Winter 2010 \& 2015 and in Summer 2016 using the Eight MIxer Receiver (EMIR).
The Fourier Transform Spectrometer (FTS) units were connected to the receivers on the 3 and 2\,mm bands, providing a resolution of 195\,kHz ($\Delta V \sim$ 0.6 and 0.4\,km\,s$^{-1}$ respectively). The Wideband Line Multiple Autocorrelator (WILMA) was connected to the higher frequency bands, providing a resolution of 2\,MHz ($\Delta V \sim$ 2.5 and 1.8\,km\,s$^{-1}$ for the 1.3 and 0.9\,mm bands respectively).
The observations were performed in wobbler switching mode with a throw of 180" (90" for the 2\,mm) in order to ensure a flat baseline.
The telescope beam size ranges from 34\arcsec\ at 72~GHz to 7\arcsec\ at 350~GHz.
Calibration uncertainties in the 3, 2, 1.3 and 0.9mm bands are typically 10, 15, 20 and 30\%, respectively. The data were reduced using the GILDAS-CLASS software\footnote{\label{gildas}http://www.iram.fr/IRAMFR/GILDAS/}. Intensities are expressed in units of antenna temperature corrected for the atmospheric absorption $T_{A}^*$. The rms noise per velocity interval of $1\kms$ expressed in units of  $T_{A}^*$ lies in the range 2--5~mK in the 3mm band, 3--7~mK in the 2mm and 1.3mm band, and 15--20~mK in the 0.9mm band. 
The spectral bands and observations properties are summarised on Table \ref{table:single-dish-recap}.

\begin{table*}
	\caption{\label{table:single-dish-recap}IRAM~30m Observational parameters.}
	\begin{tabular}{cccccccc}
		\hline\hline
		Band & Range   & Backend & Spectral Resolution  & $F_{eff}$ & $B_{eff}$ & HPBW        & rms\\
	             & (GHz)      &                & (MHz)                        &                  &         & ($\arcsec$)  & (mK, $T_A^*$)\\	\hline
		3mm & 72 -- 116  & FTS   & 0.195        &  0.95           & 0.81 -- 0.78  & 34.2 -- 21.2 & 2 -- 5  \\
		2mm & 126 -- 173 & FTS   & 0.195        &  0.94 -- 0.93    & 0.72 -- 0.64  & 19.5 -- 14.2 & 3 -- 7  \\
      1.3mm & 200 -- 310 & WILMA & 2            &  0.91 -- 0.82    & 0.59 -- 0.37  & 12.3 -- 7.9  & 3 -- 7  \\
	  0.9mm & 328 -- 350 & WILMA & 2            &  0.82           & 0.32         &    7.7 -- 7.0      & 15 -- 20 \\
		\hline
	\end{tabular}
\end{table*}

\begin{table*}
	\caption{\label{table:interf-recap} NOEMA Interferometric Observations properties.}
	\begin{tabular}{cccccccc}
		\hline\hline
Band & Range  & Backend & Spectral Resolution & Config. & Synthesized Beam              & PA      & rms \\
         & (GHz)     &               & (MHz)                        &              & ($\arcsec$ $\times$ $\arcsec$) & ($^o$) & (mJy/beam)\\
		\hline
3mm     & 85.9 -- 89.6       & WideX  &      2              &   BC   & $2.38\times2.13$        & 60       &1 -- 3  \\
		&     86.0 -- 86.2   & Narrow &      0.625          &   BC   & $2.46\times2.20$         &   52     &3  \\
		&     86.7 -- 86.9   & Narrow &      0.625          &   BC   & $2.46\times2.19$        &   48     &3  \\
		&     88.5 -- 88.7   & Narrow &      0.625          &   BC   & $2.39\times2.07$        &   -133   &3  \\
		&     89.1 -- 89.3   & Narrow &      0.625          &   BC   & $2.36\times2.06$        &   49     &3  \\
1.3mm& 216.8 -- 220.4 &WideX& 2                        &         CD       & $1.46\times 1.39$           & 72        &  3 -- 5\\
		  &   217.3 -- 217.6       & Narrow &         0.625          &        CD       &          $1.42\times1.41$            &     40      & 10\\
		  &   218.7 -- 218.9       & Narrow &         0.625          &        CD       &          $1.46\times1.37$      &      -90     & 8\\
		\hline	
	\end{tabular}
\end{table*}

\subsection{Interferometric observations}

The Cep\,E protostellar region was observed during Winter 2014 -- 2015 with the IRAM NOEMA interferometer at 3mm and 1.3mm.
At 3mm, the spectral band 85.9 -- 89.7~GHz covering the SiO $J$=2--1 line was observed in B and C configurations on 3 and 12 December 2014 and 6--7 March 2015, in a single field centered at the nominal position of the protostar $\alpha(2000)=\mathrm{23^h03^m12^s.8},\,\delta(2000)=\mathrm{61^{\circ}42'26''}$.
At 1.3mm, the spectral band 216.8--219.95~GHz covering the SiO $J$=5--4 line was observed in C and D configurations on 9 December 2014, 7 December 2015, 13 and 24 April 2015. The 1.3mm band was observed over a mosaic of 5 fields centered at offset positions ($-9\arcsec$, $-20.8\arcsec$), ($-4.5\arcsec$, $-10.4\arcsec$), ($0\arcsec$, $0\arcsec$), (+$4.5\arcsec$, $+10.4\arcsec$) and ($+4.5\arcsec$, $+20.8\arcsec$) with respect to the protostar, in order to map the emission of the protostar and the high-velocity outflow.

The Wideband Express (WideX) backend was connected to the receivers providing a resolution of 2\,MHz ($\Delta V \sim$ 2.7\,km\,s$^{-1}$ at 1.3\,mm). Narrow spectral bands with lines of interest were observed in parallel with a resolution of 256 kHz.
Observations properties are summarised on Table \ref{table:interf-recap} for both the Wide Correlator (Widex) and the Narrow correlator backends.
The antenna baselines sampled in our observations range from 19 to 176\,m at 1.3mm (from 21 to 443\,m at 3mm), allowing us to recover emission on scales from 14.7\arcsec\ to 1.5\arcsec\ at the 1.3\,mm (from 34\arcsec\ to 1.6\arcsec\ at 3mm).
At 1.3mm, for both C and D tracks, phase was stable (rms $\leq 45$\deg) and pwv was 1-3 mm with system temperatures $\sim 100$---200, leading to less than 2\% flagging in the dataset. At 3mm, phase rms was $\leq 50$\deg\, pwv was 1–10 mm, and system temperatures were $\sim 60-200\K$, leading to less than 1\% flagging in the dataset.

Calibration and data analysis were carried out following standard procedures using the GILDAS software\footref{gildas}.
The continuum emission at 1.3mm and 3mm was obtained by selecting spectral windows free of molecular line emission. Continuum-free molecular maps were subsequently produced by subtraction of the continuum contribution.

Using a natural weighting, the size of the synthesized beams at 1.3 and 3~mm are $1.46\arcsec \times 1.39\arcsec$  (PA= $72\deg$) and $3.55\arcsec \times 2.65\arcsec$ (PA=$60\deg$), respectively.
The uncertainty on the absolute flux calibration is $\leq 20\%$ and $\leq 10\%$ at 1.3 and 3~mm respectively; and the typical rms noise per spectral resolution channel is  3--5~mJy/beam and from 1--3~mJy/beam.

\section{Dust Continuum Emission}

\subsection{Source multiplicity}
\label{multiplicity}
\begin{figure}
	\includegraphics[width=0.99\columnwidth]{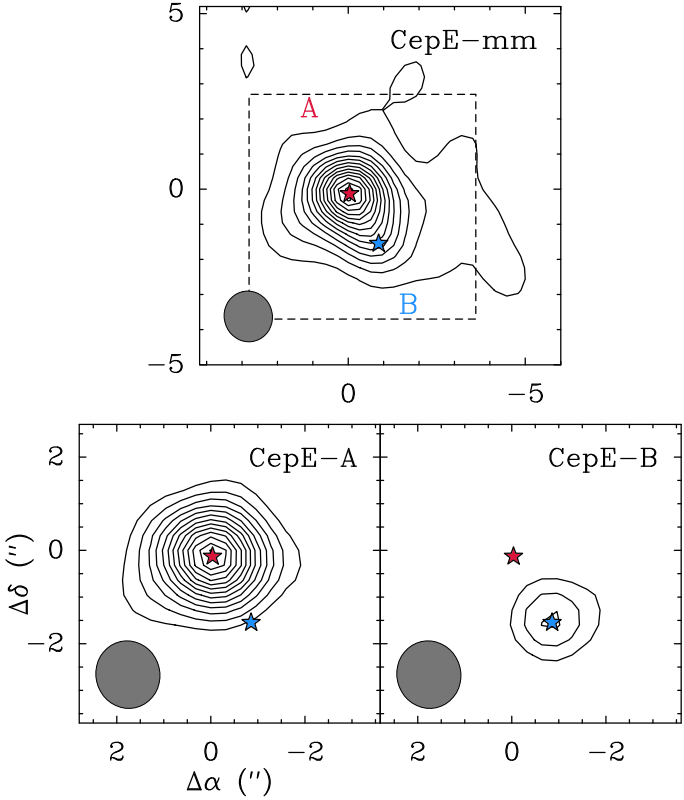}
	\caption{\label{fig:cont-A-B} \textit{Top:} Cep\,E-mm 1.3\,mm continuum map. Base contour and contour spacing are 3.6 (3$\sigma$) and 7.2\mjy/beam (6$\sigma$). The peak intensity is 99.2\mjy/1.42\arcsec-beam. The blue and red stars mark the position of Cep\,E-A and B components, respectively.
		\textit{Bottom:} Continuum emission of components A (\textit{left}) and B (\textit{right}) deduced from a 2 components fit to the visibilities (see text). Base contour and contour spacing are 10.8 (9$\sigma$) and 7.2\mjy/beam (6$\sigma$). Peak intensities are 97.1 and 26.2 \mjy/1.42\arcsec-beam, respectively.}
\end{figure}

Map of the 1.3\,mm continuum emission is displayed in the top panel of Fig.~\ref{fig:cont-A-B}. It shows a core of dimensions $1.56\arcsec \times 1.08\arcsec$  ($1141\times 788$~au; beam deconvolved), with a PA of 45~deg.
The peak of the flux distribution (99.4\,mJy/$1.4\arcsec$-beam) is located at the nominal position of Cep\,E-mm, and drives the well-known, bright, high-velocity CO outflow (Lefloch et al. 2015). Integrating over the region defined by the 3$\sigma$ contour level (3.5\mjy/1.4\arcsec-beam), the total flux of the condensation is then $S_{1.3mm}$= $233\mjy$.

The location of the peak flux is shifted with respect to the center of the continuum core, suggesting the presence of a secondary, fainter source located at about $1.7\arcsec$ southwest of the peak flux. In what follows, we will refer to the first (bright) and secondary (faint) components as Cep\,E-A and Cep\,E-B, respectively (Fig.\ref{fig:cont-A-B}).

We determined the position, size and intensity of the two continuum sources from a fit to the visibility table, assuming two 1D Gaussian distributions. We note that the uncertainties correspond to the statistical errors derived from the minimisation procedure.
We find that the bright source Cep\,E-A has a diameter of $1.20\arcsec\pm0.01\arcsec$ (875\,au) (beam-deconvolved) and a peak flux of $97.1\mjy/$1.4-beam and a total flux of $168.8\pm0.7\mjy$.
The fainter source Cep\,E-B is located at the offset position ($-0.85\arcsec$, $-1.52\arcsec$), at 1250\,au from the Cep\,E-A. It has a size of  $0.92\arcsec\pm0.02\arcsec$ (670\,au). We estimate  a peak flux of $26.2\mjy/$1.4-beam and a total flux $40.3\pm 0.6\mjy$.

In order to elucidate the nature of both condensations, we have searched for evidence of outflowing gas motions associated with Cep\,E-A and Cep\,E-B using emission maps of the SiO $J$=2--1 and 5--4 transitions present in the observed NOEMA bands, a good tracer of young  protostellar outflows (Lefloch et al. 1998).

\begin{figure}
	\includegraphics[width=0.99\columnwidth]{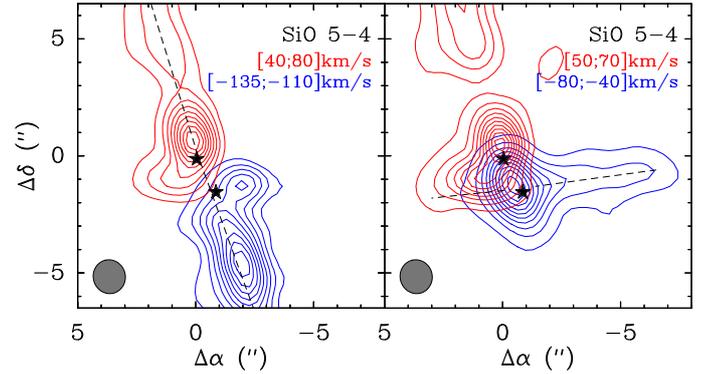}
	\caption{\label{fig:A-B-SiO} SiO $J=5-4$ jet emission. The stars points towards the fitted position of both Cep\,E-A and B cores.
		\textit{Left:} The blue and red contours represent the integrated line emission between -135 and -110 km\,s$^{-1}$ and between +40 and +80 km\,s$^{-1}$, respectively. Base contour and contour spacing of both jet emission are 10\% of their maximum intensity.
		\textit{Right:} The blue and red contours represent the  integrated line emission between -80 and -40 km\,s$^{-1}$ and between +50 and +70 km\,s$^{-1}$, respectively. Base contour and contour spacing of both jet emission are 5\% and 10\% of their maximum intensity.}
\end{figure}

The left hand panel in Fig.~\ref{fig:A-B-SiO} shows the SiO $J$=5--4 emission between $-135$ and $-110\kms$ (blue) and $+40$ and $+80\kms$ (red); SiO traces the previously known, high-velocity jet oriented in the southwest-northeast direction (Lefloch et al. 1996, 2011, 2015). Both lobes slightly overlap at the location of Cep\,E-A, confirming the association with this protostar.

The right hand panel of Fig.~\ref{fig:A-B-SiO} displays the SiO $J$=5--4 emission integrated between $-80$ and $-40\kms$ (blue) and $+50$ and $+70\kms$ (red). We note that the redshifted emission is slightly contaminated by the emission from the high-velocity jet from Cep\,E-A.
Overall, the red- and blueshifted emissions draw a collimated, jet-like structure oriented in the Eastern-Western direction with a PA of -8\deg. This jet emission is asymmetric, as it appears to be more extended along the blueshifted lobe. Both lobes overlap at about $1.5\arcsec$ South of Cep\,E-A. This definitely excludes a possible association with Cep\,E-A. Instead, the lobes appear to overlap at  the position of Cep\,E-B, supporting a physical association with the secondary dust core.

\subsection{Physical properties}

\begin{figure}
	\includegraphics[width=0.95\columnwidth]{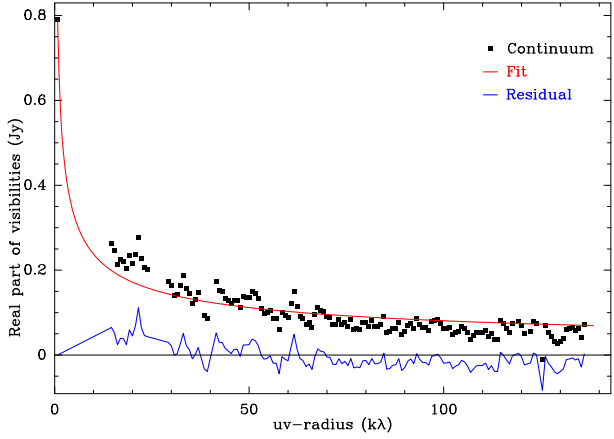}
	\caption{\label{Fig:visi_vs_base} Continuum (real part of the visibilities, black squares) and corresponding best-fit power-law model (red line) averaged over baseline bins of 1~m, as a function of baseline length. Residuals are shown as a blue line. The zero-spacing 790\mjy~flux density is extrapolated from single-dish observations at 1.3~mm by Lefloch et al. (1996).}
\end{figure}

\begin{table*}
\caption{\label{table:binary-properties}Physical properties of the Cep\,E-mm condensation and the components A and B.}
\begin{tabular}{lccccccccc}
		\hline\hline
Continuum  & Size\,\tablefootmark{a} & Size\,\tablefootmark{a} & Peak Coord. & S$^{beam}_{1.3mm}$ & S$_{1.3mm}$\,\tablefootmark{b} & $T_{dust}$\,\tablefootmark{c} &  $N(\htwo)$ & $n(\htwo)$ & M ($\msol$) \\
component	& (\arcsec) & (au) & (\arcsec) & (mJy/beam) & (mJy) & (K) & ($10^{23}\cmmd$) & ($10^7\cmmt$) & ($\msol$) \\
		\hline
Whole Core & 1.56$\times$1.08 & 1141$\times$788 & ($+0,+0$) & 99.4 & 233 & 55 & 5.7 & 4.0 & 0.83 \\
Cep\,E-A & 1.2 & 875 & ($+0,+0$) & 97.1 & 168.8 & 60 & 5.2 & 4.0 & 0.56 \\
\multirow{2}{*}{Cep\,E-B} & \multirow{2}{*}{0.9} & \multirow{2}{*}{670} & \multirow{2}{*}{($-0.8,-1.5$)} & \multirow{2}{*}{26.2} & \multirow{2}{*}{40.3} & 25 & 4.0 & 4.0 & 0.38\\
 & & & & & & 60 & 1.3 & 1.3 & 0.13\\
	\hline
	\end{tabular}
	\\
\tablefoottext{a}{FWHM, beam-deconvolved.}\\
\tablefoottext{b}{Best fit results.}\\
\tablefoottext{c}{Based on Crimier et al. (2010). For Cep\,E-B two $T_{dust}$ are assumed (see text).}\\

\end{table*}

The large-scale physical structure of Cep\,E-mm dust envelope was studied by Crimier et al. (2010). The authors derived the dust density and temperatures profiles from a 1D modelling of the dust continuum emission at 24, 70, 450, 850 and 1300\,$\mu$m by SCUBA and Spitzer. Fluxes were measured with an angular resolution in the range $7.5\arcsec$ -- $14.8\arcsec$, hence probing mainly the outer regions ($ > 8\arcsec$) of the envelope. They constrained a power law index on the density profile $n \propto r^{-p}$ with $p = 1.88$, and obtained self-consistently the temperature profile $T \propto r^{-q}$ with $q = 1.10$ in the inner region ($r < 500$~au). In the Rayleigh-Jeans approximation and for optically-thin dust emission, the emergent dust continuum emission also has a simple power-law form $I(r)\propto r^{-(p+q-1)}$.
For interferometric observations, the visibility distribution is then $ V(b)\propto b^{(p+q-3)}$, and depends only on the density and temperature power law index.

As can be seen in Fig.\ref{Fig:visi_vs_base}, a simple fit of the form $V(b) \propto b^{(-0.47 \pm 0.01)}$ reproduces well our interferometric continuum observations and the single-dish flux of Lefloch et al. (1996). We obtain a ($p+q$) value of $2.53\pm0.01$ lower than that of Crimier et al.  ($\simeq 2.98$). Such a lower value is consistent with a steeper density profile in the inner regions probed by the interferometer, and could indicate collapse of the inner layers.

In order to estimate the physical parameters of the dust condensation and the protostellar cores Cep\,E-A and B, we have adopted as dust temperature $T_{dust}$ the value estimated by Crimier et al. (2010) at the core radius ($55\K$ at $r=480$~au), and we have adopted a dust mass opacity $\kappa_{1.3mm}=0.0089$ cm$^2$g$^{-1}$
calculated by Ossenkopf \& Henning (1994), specifically their OH5 dust model which refers to grains coated by ice. Under these assumptions, we obtain a beam-averaged gas column density  N($\htwo$)= $(5.7\pm1.0)\times 10^{23}\cmmd$ in a beam of $1.4\arcsec$. From this gas column density estimate and the size of the condensation, we obtain the average gas density n($\htwo)$= $(4.0\pm0.8)\times10^{7}\cmmt$ for the condensation. This value is in agreement with the gas density derived by Crimier model $4.3 \times 10^{7}\cmmt$ at the same radius of 480~au.
From the integrated flux ($233\mjy$), we derive the mass of the condensation $M$= $0.83\pm0.17\msol$.

Our NOEMA maps show that the millimeter flux is dominated by Cep\,E-A. Under the assumption that this source dominates the Spectral Energy Distribution, we can use the Crimier model to estimate the physical parameters of
Cep\,E-A. Proceeding in the same way as above, it comes $T_{dust}\simeq$ $60\K$ at the fitted core radius $r = 440$~au. We obtain  N($\htwo$)= $(5.2\pm1.0)\times 10^{23}\cmmd$\ and $n(\htwo)$= $(4.0\pm0.8)\times 10^{7}\cmmt$ and a total dust mass $M$= $0.5\pm0.1\msol$.

For Cep\,E-B,  we constrain an upper temperature limit T$_{dust,\ max}=60$\,K similar to that of Cep\,E-A. A lower limit to $T_{dust}$ is provided adopting the temperature derived in the Crimier model at the radial distance between components A and B (1.7\arcsec)~: $T_{dust}=25\K$. Hence, we estimate the gas column density $N(\htwo)= (1.3$--$4.0)\times 10^{23}\cmmd$, the gas density  $n(\htwo)= (1.3$--$4.0)\times10^{7}\cmmt$ and the mass $M= (0.13$--$0.38)\msol$.

The physical properties of both protostars are summarized in Table~\ref{table:binary-properties}.

\section{Complex Organic Molecules}

The first hint on the presence of COMs in Cep\,E-mm was provided by Lopez-Sepulcre et al. (2015) who reported the presence of formamide (NH$_2$CHO). Following this study, we have carried out a systematic search for COM emission in the IRAM~30m survey and the NOEMA observations. We searched for emission lines from the species identified in the solar-type protostar IRAS16293-2422 (Cazaux et al. 2003; Jaber et al. 2014) and the protostellar shock L1157-B1 (Lefloch et al. 2017).  Line identification was carried out using the WEEDS package in GILDAS (Maret et al. 2011) and the public databases CDMS (Müller et al. 2005) and JPL (Pickett et al. 1998).

\subsection{Content}

\begin{figure*}
	\includegraphics[width=2.04\columnwidth]{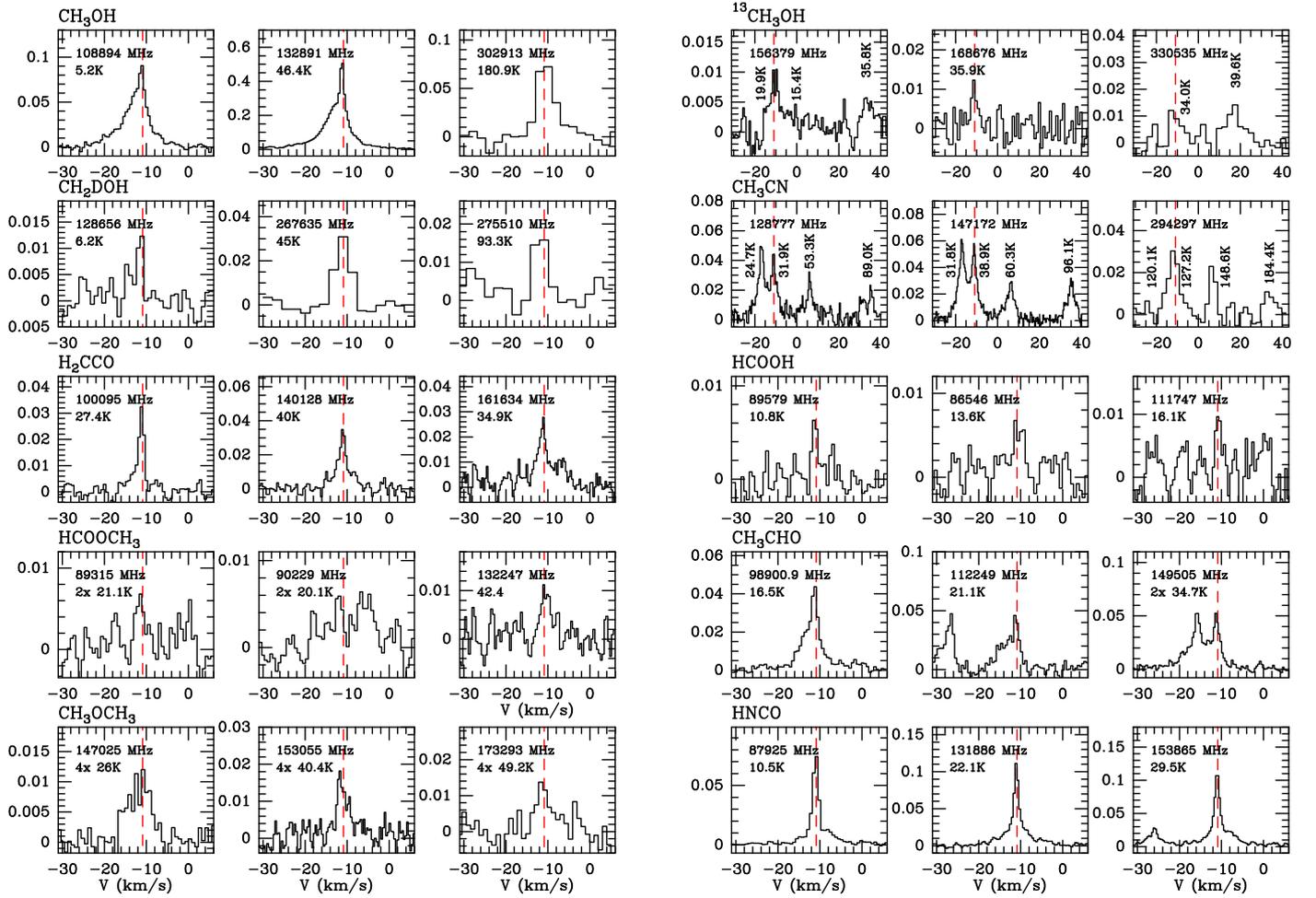}
	\caption{\label{fig:specs-montage} Montage of detected transitions from COMs and chemically related species towards Cep\,E-mm with IRAM~30m. Intensities are expressed in units of $T_A^*$. The red dashed line marks the ambient cloud velocity $v_{\mathrm{lsr}} = -10.9\kms$.}
\end{figure*}

\begin{table*}
	\caption{\label{tab:molecules-list}Molecular species (COMs and chemically related species) detected with the IRAM~30m telescope and the NOEMA interferometer. For each species, we indicate the number n of transitions detected, the corresponding range of $E_{up}$ and the detected spectral component with the 30m: extremely-broad line (eBL) component, the broad line (BL) component and the narrow line (NL) component.}
	\begin{tabular}{@{\extracolsep{4pt}}lrrrrrrrrr@{}}
	\hline\hline
					& \multicolumn{5}{c}{30m}	& \multicolumn{4}{c}{NOEMA}		\\
	\cline{2-6}\cline{7-10}
	Species	 		& n			& \multicolumn{1}{c}{$E_{up}$}	& \multicolumn{3}{c}{Components}	& n			& \multicolumn{1}{c}{$E_{up}$} & \multicolumn{2}{c}{Components}		\\
					&			& \multicolumn{1}{c}{(K)}		&	eBL	& BL	& NL	&			& \multicolumn{1}{c}{(K)}	& eBL & BL		\\
	\hline
	E-CH$_3$OH		& 72		&  4.6 -- 500.5	& $\times$&$\times$&$\times$	& 4			&  37.6 -- 500.5	&  $\times$ &  $\times$ \\
	A-CH$_3$OH		& 64		&  7.0 -- 446.5	& $\times$&$\times$&$\times$	& 2			& 373.9 -- 745.6	&  &  $\times$\\
	$^{13}$CH$_3$OH	& 21		&  6.8 -- 113.5	&		&$\times$&				& 1			& 162.4	& & $\times$		\\
	CH$_2$DOH		& 36		&  6.2 -- 123.7	& &	$\times$ & $\times$			& 4			&  10.6 -- 58.7 &  $\times$ &  $\times$	\\
	H$_2$CO			& 15		& 3.5 -- 125.8	& $\times$	& $\times$	& $\times$	& 3	& 21.0 -- 68.1 &  $\times$ & $\times$	\\
	H$_2^{13}$CO	& 9			& 10.2 -- 61.3	& $\times$	& $\times$	& $\times$	& 1	& 32.9 &  $\times$ &  $\times$\\	
	H$_2$CCO		& 18		&  9.7 -- 102.1	& & $\times$ & $\times$		& 1			& 76.5	&  $\times$ &  $\times$		\\
	HCOOH			& 6			& 10.8 -- 23.5	& & & $\times$					& 1			& 58.6	& &  $\times$		\\
	HCOOCH$_3$		& 24 		& 20.1 -- 111.5	& & $\times$ &			& 16		& 20.1 -- 111.5	&	&  $\times$	\\
	CH$_3$CHO		& 63		&  7.6 -- 89.8	& $\times$& &$\times$			& 1			& 18.3	&  &	 $\times$	\\
	CH$_3$OCH$_3$	& 30		& 11.1 -- 106.3	& & $\times$ & 					& 1			& 253.4	& & $\times$		\\
	CH$_3$COCH$_3$	& 0         &            	& 	 & &			& 1			& 119.1		& &  $\times$	\\
	HNCO			& 18		& 10.5 -- 82.3	& $\times$ & & $\times$		& 3			& 58.0 -- 231.1 &  $\times$ &   $\times$\\
	NH$_2$CHO\tablefootmark{a}		& 6			& 10.2 - 22.1	& & & $\times$	& 1	& 60.8		& &  $\times$	\\
%	HC$_3$N			& 20		& 15.7 -- 216.6	& $\times$ &$\times$ & $\times$	& 3			&  131 --452.3 &  $\times$ &  $\times$	\\
%	HC$_5$N			& 10		& 48.3 -- 104.8	& & & $\times$			& 1 		&  71.7			 & &  $\times$\\	
	CH$_3$CN		& 38		&  8.8 -- 148.6	& & $\times$ &$\times$		& * 		&   			\\
%	C$_2$H$_3$CN	& 12		& 22.1 -- 104.0	& $\times$ & & $\times$		& 0		&				\\
	C$_2$H$_5$CN	& 0 		&				& &  &				& 2		& 135.6 -- 139.9	& &  $\times$\\
	
	\hline
	\end{tabular}
	\\
	\tablefoottext{a}{30m lines from L\'opez-Sepulcre et al. 2015.}\\
	* No transition present in the band.\\

\end{table*}

The following complex O-bearing and N-bearing organic species are detected towards the Cep\,E-mm protostellar envelope:
Methanol (CH$_3$OH) and its rare isotopologues $^{13}$CH$_3$OH and CH$_2$DOH, Acetaldehyde (CH$_3$CHO), Dimethyl Ether (CH$_3$OCH$_3$), Methyl Formate (HCOOCH$_3$), Acetone (CH$_3$COCH$_3$), Formamide (NH$_2$CHO), Methyl Cyanide (CH$_3$CN) and Ethyl Cyanide (C$_2$H$_5$CN). The number of  lines detected with both instruments and the range of upper level energies $E_{up}$ (K) are given in Table~\ref{tab:molecules-list}. A montage of illustrative transitions from each species observed with the IRAM~30m telescope  is displayed in Fig. \ref{fig:specs-montage}.
In this work, we also report the detection of the following species, chemically related to COMs:
Formaldehyde (H$_2$CO), Ketene (H$_2$CCO), Formic acid (HCOOH), Isocyanic Acid (HNCO).

Because of the limited spectral coverage of NOEMA observations and the lower sensitivity of the IRAM~30m survey, there is only a partial match between the sets of detected lines detected with both instruments. We note that, apart from CH$_3$CN (not observed with NOEMA), a)~Propanone (CH$_3$COCH$_3$) and Ethyl Cyanide (C$_2$H$_5$CN) were not detected by the IRAM~30m telescope; b)~Vinyl cyanide (C$_2$H$_3$CN) was not detected by NOEMA.
Our NOEMA observations show bright emission from highly excited molecular transitions, like
HCOOCH$_3$ 20$_{0,20}$--19$_{0,17}$ ($E_{up}=111.5\K$) or CH$_3$OH-A $6_{1,-1}$--$7_{2,-1}$  ($E_{up}$=$373.9\K$) which are not detected with the IRAM~30m. This effect results from the dilution of the emission in the telescope main beam.

\begin{figure}
	\center
	\includegraphics[width=1\columnwidth]{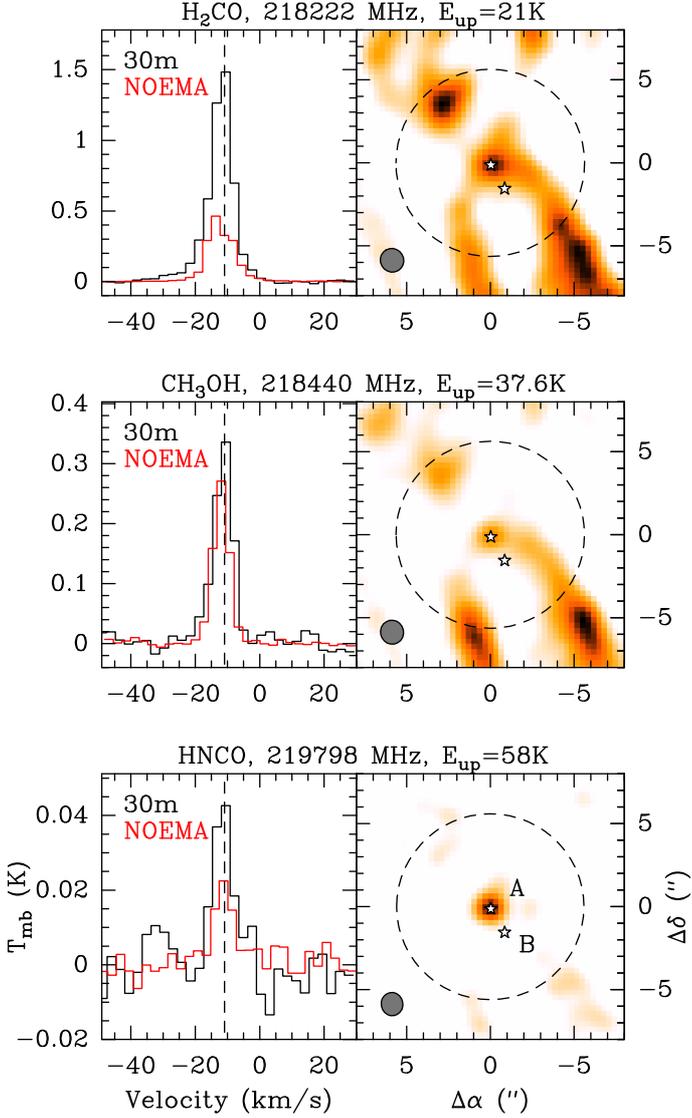}
	\caption{\label{fig:30m+NOEMA_comp} H$_2$CO 3$_{0,3}$ -- 2$_{0,2}$, CH$_3$OH 4$_{2,0}$ -- 3$_{1,0}$  and HNCO 10$_{0,10}$ -- 9$_{0,9}$  transitions detected with both IRAM~30m and NOEMA towards Cep\,E-mm. \textit{Left:} 30m spectra (black) and NOEMA spectra convolved to the single-dish beam (red). Both spectra are displayed with the same spectral resolution. The black dashed line marks the cloud velocity  $v_{\mathrm{lsr}} = -10.9\kms.$ \textit{Right:} emission map obtained with NOEMA. The interferometric synthesized beam is represented by the grey disc. White star marks the observing position of the 30m and the beam width at half-power is drawn by a black dashed circle. The locations of protostars Cep\,E-A and Cep\,E-B are marked by stars.}
\end{figure}

In order to estimate the amount of flux filtered out by NOEMA, we compared the fluxes of molecular transitions detected at $\geqslant 5\sigma$. The results obtained are summarized in Table~\ref{lineflux}.
We found that all the flux is recovered for lines from high $E_{up}$, like CH$_3$OH 8$_{0,0}$ -- 7$_{1,0}$ ($E_{up}$= $88.7\K$). For the other transitions of lower excitation, the interferometer recovered from 20\% to 60\% of the flux collected with the IRAM~30m. This is illustrated in  the case of H$_2$CO $J$= $3_{0,3}$--$2_{0,2}$ ($E_{up}$= $21\K$), CH$_3$OH $J$= $4_{2,0}$--$3_{1,0}$ ($E_{up}$= $45.5\K$) and HNCO 10$_{0,10}$--9$_{0,9}$ ($E_{up}$= $58.0\K$) in Fig.~\ref{fig:30m+NOEMA_comp}.

\begin{table}
\caption{\label{lineflux} Fluxes of molecular transitions detected at $\geqslant 5\sigma$ recovered by NOEMA and compared to the 30m telescope.}
\begin{tabular}{lrrrrrr}
	\hline\hline
	Species		& Frequency	& Quantum				& $E_{up}$	 & Recov.	\\
				& (MHz)		& Numbers				& (K)		&   (\%) 		\\ \hline
	CH$_3$OH 	& 218440	&4$_{2,0}$ -- 3$_{1,0}$	& 37.6		&  60		\\
				& 216946	&5$_{1,0}$ -- 4$_{2,0}$	& 48.0		&  50		\\
				& 220079	&8$_{0,0}$ -- 7$_{1,0}$	& 88.7		&  90		\\
	CH$_2$DOH	& 86669		&2$_{1,1,0}$ -- 2$_{0,2,0}$	& 10.6	&      20		\\
	H$_2$CO		& 218222	&3$_{0,3}$ -- 2$_{0,2}$	& 21.0		&  30	 \\
				& 218760	&3$_{2,1}$ -- 2$_{2,0}$	& 68.1		&  60  \\
	HNCO		& 219798	& 10$_{0,10}$ -- 9$_{0,9}$	& 58.0	&  50		\\
   	\hline
\end{tabular}
\end{table}

\subsection{Spectral signatures}
\begin{figure}
	\includegraphics[width=0.98\columnwidth]{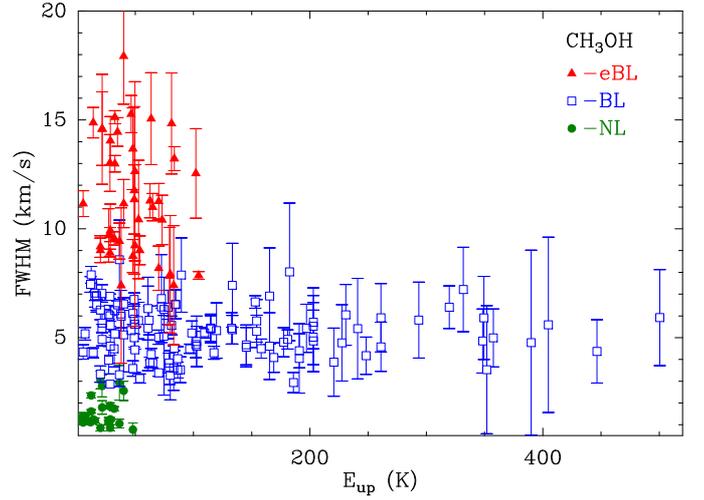}
	\caption{\label{fig:meth-fwhm-up} Linewidth of the three physical components identified in the CH$_3$OH transitions as a function of the $E_{up}$.}
\end{figure}

\begin{figure}
	\includegraphics[width=0.98\columnwidth]{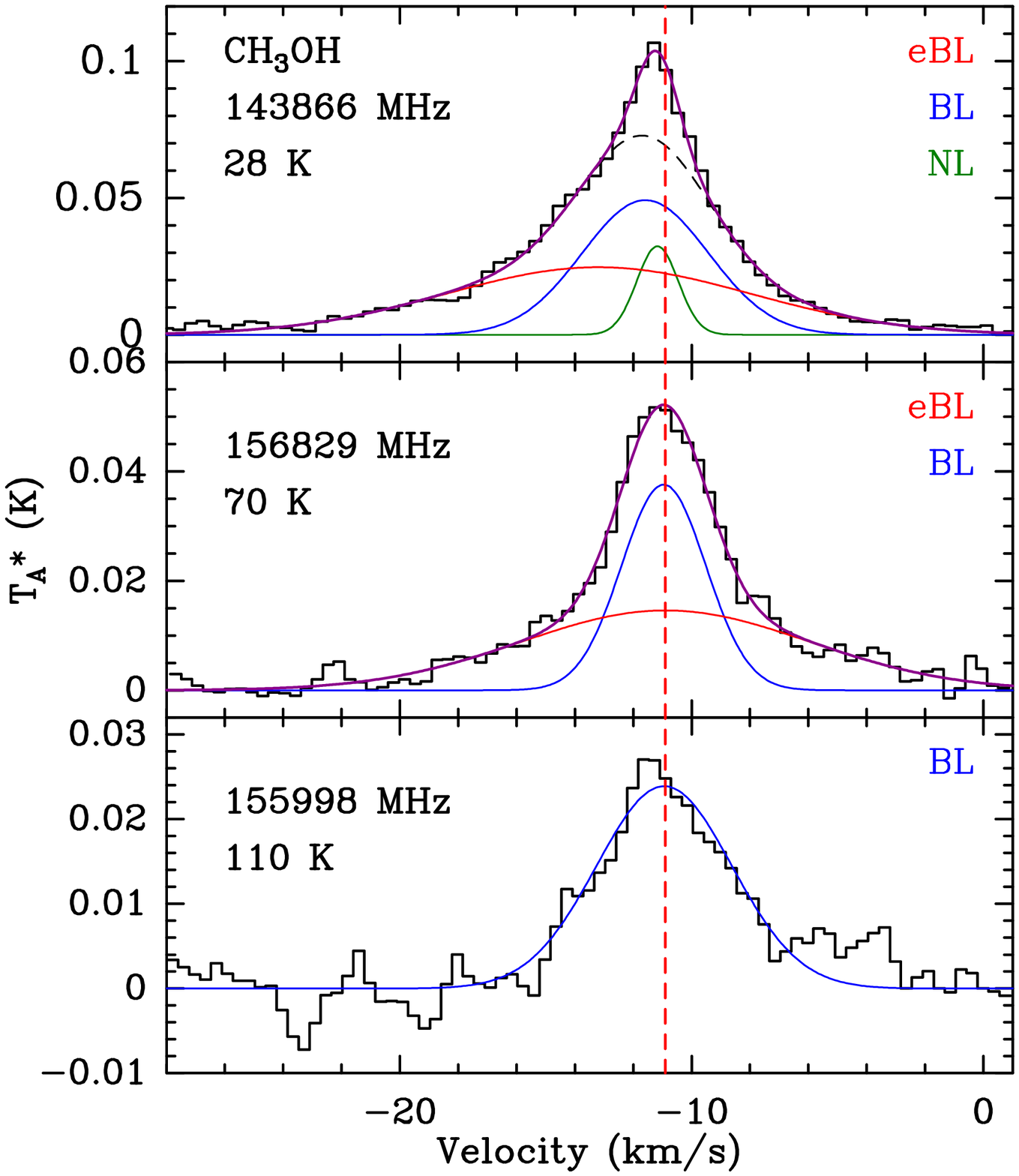}
	\caption{\label{fig:meth-3comp-prof} Methanol line profiles as observed with IRAM~30m with an angular resolution of 15.8\arcsec. The ambient cloud velocity ($v_{\mathrm{lsr}} = -10.9\kms$) is marked by the dashed red line.
	\textit{Top:} A-CH$_3$OH $J$=3$_{1,+0}$--2$_{1,+0}$ ($E_{up}=28.3\K$) transition. Three Gaussian components are fitted to the line profile: the eBL in red, the BL in blue and the NL in green. Dashed black curve shows the sum of eBL and BL components. In purple the overall fit.
	\textit{Middle:} E-CH$_3$OH $J$=7$_{0,0}$--7$_{-1,0}$ ($E_{up} = 70.2\K$) transition. Only the eBL (red) and BL (blue) components are detected.
	\textit{Bottom:} E-CH$_3$OH $J$=9$_{0,0}$--9$_{-1,0}$ ($E_{up} = 109.6\K$) transition. Only the BL (blue) component is detected.}
\end{figure}

Figure~\ref{fig:specs-montage} shows that a large variety of profiles is observed, depending on the molecular species and the line excitation conditions. For instance, narrow line components ($\simeq 1-3\kms$) peaking at the cloud systemic velocity  ($v_{lsr}$=$-10.9\kms$) are detected in  HCOOH and CH$_3$CHO. Broad velocity components extending over $20\kms$ are also detected in the profiles of H$_2$CCO, or CH$_3$CHO.
We have identified three types of spectral line features, depending on the linewidth, the velocity range and the emission peak velocity: a narrow line component (NL), a broad line component (BL) and the extremely-broad line component (eBL).

The three components are detected in the low-excitation transitions of methanol, with
$E_{up}$ < $50\K$ . The eBL and BL components are both found in transitions up to $E_{up} < 100\K$, whereas only the BL component is detected higher excitation transitions with $E_{up} > 100\K$.
This is shown in Fig.~\ref{fig:meth-fwhm-up} in which we have reported the linewidth of each component as a function of the upper energy level $E_{up}$ for all methanol transitions, as derived from a gauss fitting to the line profiles.
Figure~\ref{fig:meth-3comp-prof} shows three methanol transitions with $E_{up} = 28\K$, 70\K\ and 110\K\ respectively, observed in the 2mm band at approximately the same angular resolution ($15.8\arcsec$). It comes out that the low- and mid-excitation line methanol emission ($E_{up}$= $28\K$ and $70\K$) is dominated by the contribution of the extremely-broad line component, whereas the high-excitation line emission is dominated by the BL component.

We have applied the same decomposition to the line profiles of the other detected molecular species that are listed in Table~\ref{tab:molecules-list}. We describe their properties in more detail hereafter.

\subsubsection{The extremely-broad line (eBL) component}
Methanol spectra show a extremely-broad line component with a line width (FWHM) $\Delta v \geqslant 7 \kms$ (Fig.~\ref{fig:meth-fwhm-up}) and peaking at velocities up to $\pm 10\kms$ with respect to the source ($v_{lsr}$=$-10.9\kms$). The line emission can be reasonably well fitted by a gaussian centered at a velocity shifted by 1--$3\kms$ from the source (Fig.~\ref{fig:meth-3comp-prof}, top and middle panel; also Fig.~\ref{fig:specs-montage}). It is detected in the transitions with $E_{up}\leq 100\K$, only.
This component is detected with NOEMA and traces the cavity walls of the high-velocity outflow driven by Cep\,E-A. We estimate a typical size of $20\arcsec$ (see Lefloch et al. 1996; 2015).
This component is also detected in the lines of formaldehyde, methyl cyanide and acetaldehyde (see Table~\ref{tab:molecules-list}).

\subsubsection{The narrow line (NL) component}
A narrow line component with $\Delta v \leqslant 3 \kms$ peaking at the systemic velocity of the cloud, is detected only in the profiles of the low-excitation CH$_3$OH lines ($E_{up}\leq 50\K$). This component is present in the line spectra of many COMs (CH$_3$CHO, HCOOH, CH$_3$CN, and H$_2$CCO) with $E_{up} \leq 60\K$, when observed with the IRAM~30m.
The flux filtered by the interferometer is about 60--80\% for these transitions (see Table~\ref{lineflux}). We note that the low velocity dispersion and the low excitation conditions of this gas component are consistent with an origin in the cold and quiescent outer envelope.

\subsubsection{The broad line (BL) component}

The BL component is seen in all the methanol transitions detected with IRAM 30m.
This component differs from the bipolar outflow as a)~its emission peaks at the source systemic velocity, b)~its linewidth varies little with $E_{up}$, c)~it is detected in the high-excitation lines of methanol, with $E_{up}$ in the range 100--$500\K$ (see Tables~A.1--A.2). We find a very good match between the high-excitation line profiles of methanol observed with the NOEMA and those observed with the IRAM~30m. This confirms that we are actually detecting the same component. Interestingly, the line profiles of COMs, as observed with NOEMA, can be well fitted by a gaussian with a line width of $\simeq 5\kms$, similar to that of the high-excitation CH$_3$OH lines.

\subsection{Spatial distribution}

\begin{figure*}
	\includegraphics[width=2\columnwidth]{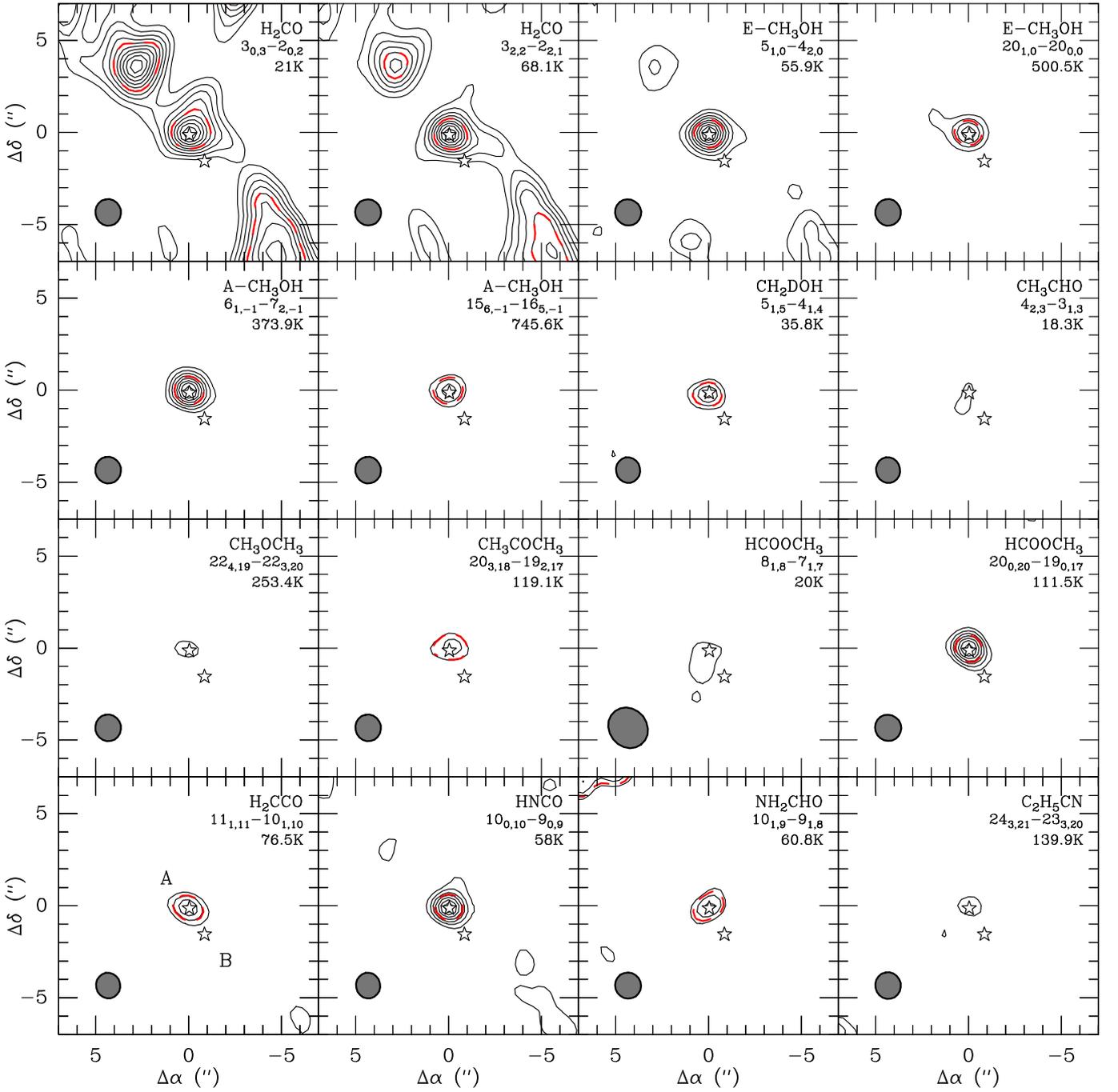}
	\caption{\label{fig:cartes} Integrated emission maps of COMs and chemically related species. For  H$_2$CO, base contour and contour spacing are 10\% of the maximum integrated intensity. For the E-CH$_3$OH map base contour and contour spacing are 3$\sigma$. For the others molecules, base contour and contour spacing are 3 and $2\sigma$ of each map. Red-dashed lines corresponds to half maximum intensity if it is greater than 3$\sigma$ noise. The locations of Cep\,E-A and Cep\,E-B are marked by stars.}
\end{figure*}

We have produced emission maps of selected molecular transitions of COMs and chemically related species from the calibrated UV table adopting a natural weight. A montage of such maps is shown in Fig.~\ref{fig:cartes}. The emission of the various COMs appears to be compact and centered on the dust emission peak of Cep\,E-A. Interestingly, there is no molecular emission centered on Cep\,E-B, the second core component.

We have estimated the size of the COM emission region around Cep\,E-A for the following bright molecular lines: CH$_3$OH 5$_{1,0}$ -- 4$_{2,0}$  and 6$_{1,5}$ -- 7$_{2,6}$, CH$_2$DOH 5$_{1,5}$ -- 4$_{1,4}$  and HCOOCH$_3$ 20$_{0,20}$ -- 19$_{0,19}$.
We performed 1D Gaussian fits to the visibilities in the UV plane around the Cep\,E-A. The result of the procedure does not appear to depend significantly on the $E_{up}$ of the transition, which ranges between $36\K$ and $112\K$. We obtained a typical size  of $0.7\arcsec \pm 0.1\arcsec$ (FWHP) for all the transitions, i.e. the emission region is marginally resolved.

As can be seen in Fig.~\ref{fig:cartes}, the emission of  H$_2$CO 3$_{2,2}$--2$_{2,1}$ ($E_{up} = 68\K$) appears slightly more extended than that of CH$_2$DOH and other COMs. A fit to the  H$_2$CO 3$_{2,2}$--2$_{2,1}$ distribution yields a size of $1.2\arcsec$. For COMs like CH$_3$OCH$_3$, CH$_3$COCH$_3$ and H$_2$CCO 3$_{0,3}$--2$_{0,2}$, the intensity of the detected lines is much lower and the results of a UV visibility fitting procedure are too uncertain. In such cases, we estimated the size of the emitting region directly from a simple Gaussian fit to integrated intensity maps, deconvolved from the synthesized beam ($1.4\arcsec$). The results display relatively little scatter, and we find sizes ranging between $0.6\arcsec$ and $0.8\arcsec$.

We note that the emission of the low $E_{up}$ transitions of CH$_3$OH, H$_2$CO and HNCO also trace an extended component associated with the cavity walls of the outflow system (Lefloch et al. 2015).

To summarize, our NOEMA observations  provide  direct evidence for a hot corino region around Cep\,E-A. Estimates of the emitting regions of various COMs yield a typical size of about $0.7\arcsec$, with little scatter. This size is close to the half power beamwidth of the synthetic beam, so that the hot corino region is only marginally resolved.

\subsection{Molecular abundances in Cep\,E-A}
\label{Section:ROTD}

\begin{table}
\caption[]{\label{tab:rotd-30m}Physical properties of the molecular emission detected towards Cep\,E-A: rotational temperature, column density and molecular abundances relative to \htwo. An \htwo\ column density of $4.0 \times 10^{23}\cmmd$ was adopted (see text). A source size of 1.2\arcsec and 0.7\arcsec\ was assumed for H$_2^{13}$CO and the other COMs emission, respectively.}
\begin{tabular}{@{\extracolsep{1pt}}lrrr@{}}
\hline\hline
Species				& T$_{rot}$		& $N$				&  $X$				\\
					& (K)			& ($10^{15}\cmmd$)	&	($10^{-9}$)		\\
\hline
$^{13}$CH$_3$OH 	& $27.3\pm5.3$	& $7.0\pm2.2$		&	$17.5\pm5.5$	\\
CH$_2$DOH			& $20.2\pm1.2$	& $17.7\pm2.5$		&	$44.3\pm6.3$	\\
H$_2^{13}$CO		& $14.0\pm2.4$	& $0.5\pm0.2$		&	$1.3\pm0.5$		\\
H$_2$CCO			& $19.6\pm1.2$	& $3.2\pm0.5$		&	$8.0\pm1.3$		\\
CH$_3$OCH$_3$		& $37.9\pm1.9$	& $14.4\pm1.5$		&	$36.0\pm3.8$	\\
HCOOCH$_3$			& $29.8\pm6.8$	& $10.2\pm2.5$		& 	$25.5\pm6.2$	\\
HCOOH				& 30			& $0.32\pm0.12$		&	$0.8\pm0.3$		\\
					& 60			& $0.34\pm0.13$		&	$0.9\pm0.3$		\\
CH$_3$CHO			& 30			& $4.1\pm1.0$		&	$10.3\pm2.5$	\\
					& 60			& $5.0\pm1.2$		&	$12.5\pm3.0$	\\
CH$_3$COCH$_3$		& 30			& $3.3\pm0.8$		&	$8.3\pm2.0$		\\
					& 60			& $1.0\pm0.2$		&	$2.5\pm0.5$		\\
HNCO				& 30			& $0.64\pm0.15$		&	$1.6\pm0.4$		\\
					& 60			& $0.64\pm0.15$		&	$1.6\pm0.4$		\\
NH$_2$CHO			& 30			& $0.18\pm0.04$		&	$0.5\pm0.1$		\\
					& 60			& $0.18\pm0.04$		&	$0.5\pm0.1$		\\
CH$_3$CN			& $32.2\pm4.4$	& $1.4\pm0.5$		&	$3.5\pm1.2$		\\
C$_2$H$_5$CN		& 30			& $0.42\pm0.18$		&	$1.1\pm0.5$		\\
					& 60			& $0.18\pm0.07$		&	$0.5\pm0.2$		\\
\hline
\end{tabular}
\\
\end{table}

In this section, we discuss the physical properties (excitation temperature, column densities) of the COMs and the chemically related species detected towards the hot corino of Cep\,E-A.
The hot corino properties were obtained from a population diagram analysis of their Spectral Line Energy Distribution (SLED) or under the assumption of Local Thermodynamical Equilibrium (LTE), when only one line was detected, as is the case for CH$_3$CHO, CH$_3$COCH$_3$, HCOOH, NH$_2$CHO, C$_2$H$_5$CN and HNCO.
The source size, as determined in Section 4.2, was taken into account and the derived column densities are source-averaged. In our population diagram analysis, line opacities are taken into account following the approach of Goldsmith \& Langer (1999). Under the LTE assumption, in the absence of constraints on the excitation temperature, calculations were done for the following excitation temperatures $T_{ex}$: 30\K\ and $60\K$. The first value is similar to the determinations of $T_{rot}$ of COMs obtained when a population diagram analysis was feasible (see below; also Table~6), while the second value is consistent with the gas kinetic temperature predicted at the hot corino radius  by Crimier et al. (2010; see also below Sect.5.1). In that way, we obtain a plausible range of values for  molecular column densities.

For the molecular species detected with NOEMA, the SLED was directly measured from the WideX spectra at the location of Cep\,E-A.
For the IRAM~30m data, we first applied the spectral decomposition discussed in Sect.~4.2 in order to obtain the flux of the three physical components (bipolar outflow, cold envelope, hot corino) and  then built the SLED of the hot corino. The list of all the identified transitions, along with their spectroscopic and observational properties are given in Appendix A.
We have adopted a hot corino size of $0.7\arcsec$ for  all the COMs and of  1.2\arcsec\ for H$_2^{13}$CO, in agreement with our NOEMA observations.

\begin{figure*}
\includegraphics[width=1.9\columnwidth]{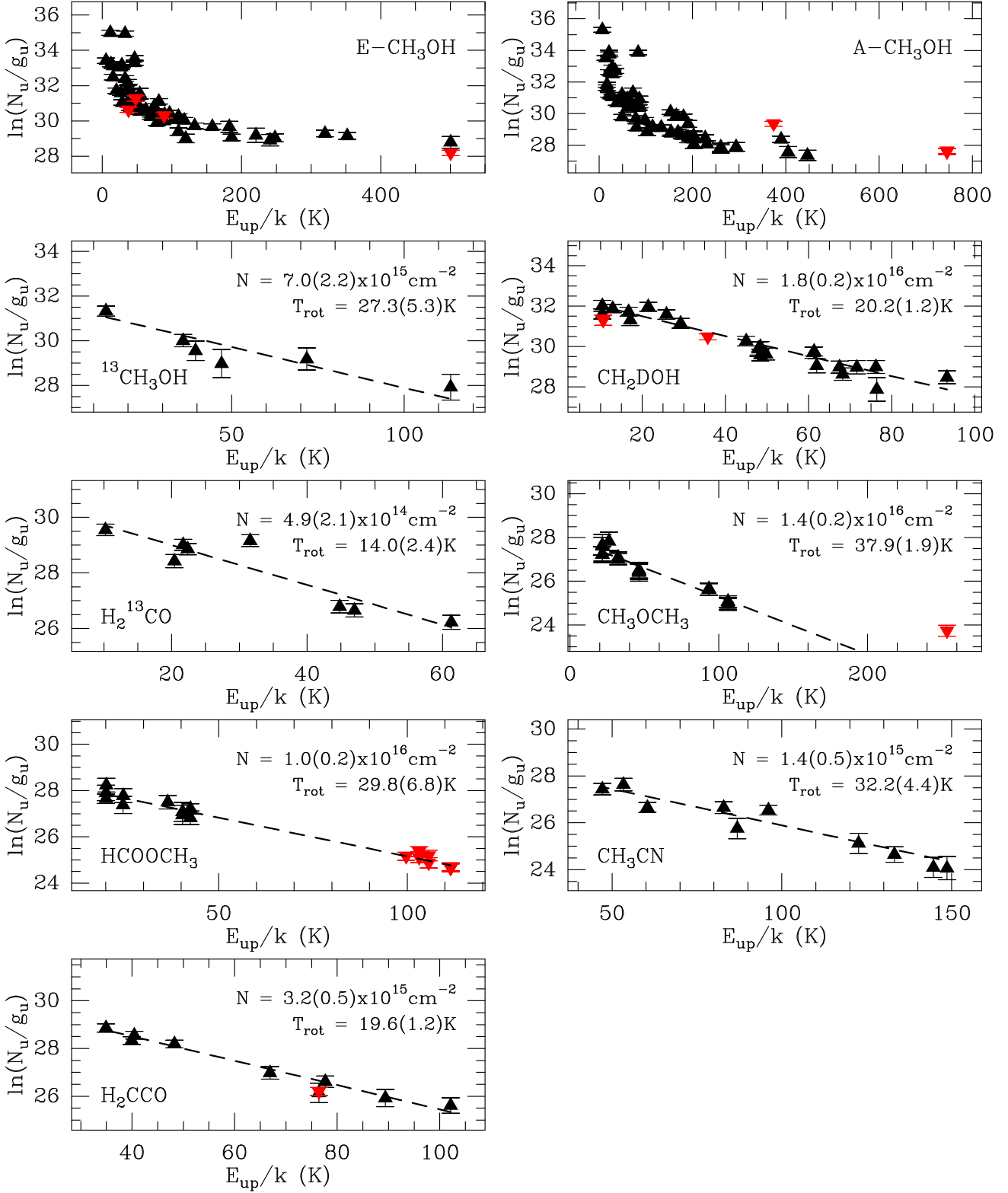}
\caption{\label{fig:ROTD_fig} Population diagram analysis of the COMs detected in the hot corino of Cep\,E-A.  A size of $0.7\arcsec$ was assumed, except for H$_2^{13}$CO ($1.2\arcsec$). Black (red) triangles represent the IRAM~30m (NOEMA) fluxes. The best fitting solution using only the 30m fluxes is drawn by a black dashed line.}
\end{figure*}

The rotational diagrams for E-/A-CH$_3$OH, $^{13}$CH$_3$OH, CH$_2$DOH, H$_2^{13}$CO, CH$_3$OCH$_3$, CH$_3$CN are presented in Fig.~\ref{fig:ROTD_fig}. We have superimposed the flux obtained with NOEMA.  Molecular abundances were derived adopting a total \htwo\ column density N(\htwo) = $4.0 \times 10^{23} \cmmd$ based on our continuum emission analysis (see Sect.~3.2). The results of our analysis (rotational temperature, source-averaged column density and molecular abundances relative to $\htwo$) are presented in Table~\ref{tab:rotd-30m}.

\subsubsection{Methanol}

The population diagrams for A-/E-CH$_3$OH are displayed in Fig.~\ref{fig:ROTD_fig}. They deviates strongly from linearity so that the SLED cannot be fitted by a single rotational temperature. The presence of excitation gradients in the source, line opacity effects and radiative pumping (see e.g. Leurini et al. 2016) could account for this behaviour.
We have estimated the opacity of a few CH$_3$OH transitions detected in the IRAM~30m survey. We compared the main-beam temperatures of two sets of CH$_3$OH and $^{13}$CH$_3$OH transitions of similar excitation, $3_{0,+}$--$2_{0,+}$ ($E_{up}=14\K$) and $5_{2,-}$--$4_{2,-}$ ($E_{up}=73\K$) adopting a typical $^{12}$C/$^{13}$C abundance ratio of 70 (Langer et al. 1992) and under the standard approximation that both molecules share a similar excitation temperature. It comes out $\tau^{12} \simeq 3$ (20) for the $3_{0,+}$--$2_{0,+}$ ($5_{2,-}$--$4_{2,-}$) transition.
We conclude that the CH$_3$OH emission is optically thick over a wide range of excitation conditions.
Unlike the main isotopologue, the rotational diagrams of CH$_2$DOH and $^{13}$CH$_3$OH are well fitted by a single temperature component.

The population diagram analysis of $^{13}$CH$_3$OH yields to N($^{13}$CH$_3$OH)= $(7.0\pm 2.2)\times 10^{15}\cmmd$ and $T_{rot}$= $(27.3\pm 5.3)\K$. The opacities of the detected $^{13}$CH3OH transitions were computed under LTE conditions, and were all found to be $\tau^{13} \leqslant 0.2$, therefore optically thin. As a consequence, we used the optically thin $^{13}$CH$_3$OH emission to determine the total methanol column density. Assuming a $^{12}$C/$^{13}$C abundance ratio of 70, we obtain the total methanol column density N(CH$_3$OH)=  $(4.9 \pm 1.5)\times 10^{17}\cmmd$.

Six and eight transitions of A- and E-CH$_3$OH, respectively, with $E_{up} > 200\K$ were detected in the IRAM survey. The flux of these transitions cannot be accounted for by the component detected in the $^{13}$CH$_3$OH lines, and requires higher excitation conditions. A simple population diagram analysis in the range $E_{up}$= 150--500\K, yields $T_{rot}=440 \pm 210\K$, $N=(1.3\pm0.4)\times 10^{16}\cmmd$ and $T_{rot}=460 \pm 220\K$, $N=(4.2\pm1.1)\times 10^{16}\cmmd$ for A- and E-species, respectively.
One might note a discrepancy between the column densities of E- and A-CH$_3$OH. However, taking into account the uncertainties on the  derived column densities,  both estimates are consistent within the $2\sigma$ level.
The total CH$_3$OH column density is $N\sim 5\times 10^{16}\cmmd$, which is one order of magnitude less than the amount of CH$_3$OH traced by the rare isotopologues. Observations at high-angular resolution and better sensitivity would help clarifying the nature of this component.

\subsubsection{Other COMs and chemically related species}
The population diagrams for the other COMs  and some chemically related species(H$_2^{13}$CO, H$_2$CCO, CH$_3$OCHO,CH$_3$OCH$_3$, CH$_3$CN and HC$_3$N) are presented in Fig.~\ref{fig:ROTD_fig}. As for $^{13}$CH$_3$OH, the population diagrams are reasonably well fitted by a straight line.

As can be seen in Table 6, All COMs display rather similar rotational temperatures in the range $T_{rot}$= 20--$40\K$. No trend is observed as a function of the rotational dipole moment, which is not unusual (see e.g. Lefloch  et al. 2017).
The H$_2$CO physical properties were estimated from its $^{13}$C isotopologue following the same procedure as for CH$_3$OH. We obtain the total formaldehyde column density N(H$_2$CO)= $(3.5\pm1.4)\times 10^{16}\cmmd$.

Because of the limited spectral coverage of the NOEMA observations, we could only detect one transition from a few COMs (see Table 4). As explained above, the column density and the molecular abundance were estimated for $T_{ex}$= $30\K$ and $T_{ex}$= $60\K$.
Whereas the first value is similar to the $T_{rot}$ values of COMs obtained from a population diagram analysis (see below; also Table~6), the second value is consistent with the gas kinetic temperature predicted at the hot corino radius by Crimier et al. (2010; see also below Sect.~5.1). This procedure allows us to derive a range of column densities along with their uncertainties.
As for HCOOH, CH$_3$CHO, HCNO, NH$_2$CHO, the results do not vary significantly with the value of $T_{ex}$ adopted, and the differences lie within the statistical uncertainties. For these species, we have reported as abundance in Table 8, the average results between the $30\K$ and $60\K$ values.
On the contrary, in the case of CH$_3$COCH$_3$ and C$_2$H$_5$CN, large differences, up to a factor of 3, are found when $T_{ex}$ varies from $30\K$ to $60\K$. For these species, we have  reported the range of obtained values in Table 8. We have also reported in Table~9 the molecular abundances of COMs relative to that of HCOOCH$_3$, a molecule which is often detected in young stellar objects (see e.g. Lefloch et al. 2018).

The molecular column densities of O-bearing species are found relatively similar, of the order of a few $10^{15}\cmmd$, except for HCOOH which is less abundant by one order of magnitude. By comparison, N-bearing species (CH$_3$CN, C$_2$H$_5$CN) are less abundant with column densities lower than $10^{15}\cmmd$. We note that NH$_2$CHO and HNCO, which both display a peptide bond NH--C=O, harbour abundances similar to N-bearing species.

To summarize, a simple population diagram analysis towards Cep\,E-A reveals a rich and abundant content in COMs, typical of hot corinos. This is the first detection of a hot corino around an isolated intermediate-mass protostar.

\section{Discussion}
\subsection{Physical structure of Cep\,E-mm}

Our observations with NOEMA have shown evidence for a system of 2 protostars in the dust condensation associated with Cep\,E-mm. This is consistent with the work by Moro-Martin et al. (2001), who reported  hints of multiplicity in  Cep\,E-mm. These authors detected two sources in the dust core observed at 222\ghz, whose coordinates are in good agreement with those of Cep\,E-A and Cep\,E-B.
They derived the physical parameters of both sources assuming a lower dust temperature (18\K) and dust mass  opacity ($\kappa=0.005\cmmd\gmu$) than the values adopted here. This results into a higher envelope mass $13.6\msol$ and higher cores masses of 2.5 and $1.8\msol$ for Cep\,E-A and Cep\,E-B, respectively.
Taking into account their hypothesis on $T_{dust}$ and $\kappa$, our mass estimates are in good agreement with theirs.

As discussed in Sect.~3.2, we have adopted a dust temperature $T_{dust}=55\K$ and dust mass opacity $\kappa_{1.3mm} = 0.0089\cmmd\gmu$ in a region of 950 au ($1.3\arcsec$), corresponding to the mean size of the Cep\,E-mm condensation. We derived a mean gas density $n(\htwo)=(4.0\pm0.8)\times10^{7}\cmmt$ which is in excellent agreement with the value predicted by Crimier et al. (2010).
Our continuum observations are therefore consistent with the presence of warm dust and gas at $0.7\arcsec$ scale around Cep\,E-A. However, NOEMA reveals CepE-mm as a binary system and the spherical symmetry hypothesis is no longer valid at $1.7\arcsec$ scale, the separation between both components (Fig~\ref{fig:cont-A-B}).

\subsection{Chemical differentiation}

Our COMs emission maps show that the flux distributions are strongly peaked towards  Cep\,E-A, while there is barely any  molecular emission detected towards Cep\,E-B (Fig. \ref{fig:cartes}).
Many of the COM transitions detected towards Cep\,E-A are rather weak and are sometimes marginally detected (see Table A.6). Since component B displays a smaller size and a lower gas column density, it is worth investigating whether the apparent chemical differentiation could be biased by the sensitivity of the data.
We have applied a simple scaling to the intensities of the COM transitions detected towards Cep\,E-A, taking into account the difference of the continuum source size and peak flux. This provides a reasonable approximation to the line fluxes towards Cep\,E-B assuming that the rotational temperature and the physical conditions are the same.
It comes out that 5 transitions of methanol and 2 of HNCO should be measured above $5\sigma$ noise level but only the brightest methanol transition is detected towards Cep\,E-B.
For all others COMs, the expected intensity then fall under the $3\sigma$ detection limit. Only the lack of CH$_3$OH  towards Cep\,E-B provides evidence for a different chemical composition in COMs.
Recently, very sensitive observations of Cep\,E-mm have been obtained as part of the NOEMA Large Program "Seeds Of Life In Space" (SOLIS; Ceccarelli et al. 2017) and confirm the chemical differentiation observed between protostars Cep\,E-A and B (Lefloch 2018 in prep).

Such a chemical differentiation has been reported in other multiple systems, like the low-mass protostars IRAS16293-2422 (Bottinelli et al. 2004a; J{\o}rgensen et al. 2011, 2016), IRAS4A (Santangelo et al. 2015; L\'opez-Sepulcre et al. 2017), and, recently, towards the intermediate-mass protostars NGC2264 CMM3 (Watanabe et al. 2017). With four examples at hand, we speculate that it could be a general feature of multiple protostellar systems and not a "pathological anomaly".
There is no systematic trend between the millimeter thermal dust and molecular line emission. Towards IRAS16293-2422 and IRAS4A, a rich content in COMs is observed towards the source with the less massive continuum source. Towards NGC2264 CMM3, it is the most massive continuum component which displays a rich molecular content. The case of Cep\,E-mm appears similar to the latter one.

High-mass star forming regions (HMSFRs) also present a rich chemical diversity.  One of the best known examples is provided by  Orion-KL. This source harbours: \textit{i)} a dichotomy between the spatial distribution of complex O-bearing and complex nitrogen bearing species, with the latter species probing the hotter gas (see, e.g. Guelin et al. 2008; Favre et al. 2011; Friedel et al. 2012; Peng et al. 2013; Brouillet et al. 2013; Crockett et al. 2014, 2015) but also, \textit{ii)} differences between supposed chemically related species (see, e.g. Favre et al. 2017; Pagani et al. 2017).
Other examples are provided  by W3(OH)(Qin et al. 2015; Nishimura et al. 2017)  and SgrB2 (Belloche et al. 2008, 2013).
In an observational study of four HMSFRs (Orion KL, G29.96, IRAS 23151+5912, IRAS
05358+3543), Beuther et al. (2009) showed that the properties of CH$_3$OH can be easily accounted for by the physical conditions  (temperature) in the cores, whereas the  N-bearing species appear to be more selective as they are detected only towards the sources at the (evolved) hot core stage. Recently, in  an ALMA study of the filamentary HMSFR G35.20, S\'anchez-Monge et al. (2014) found that only three out of the six continuum cores of the filament display COM emission typical of hot cores.

Several hypotheses have been proposed to account for the observed chemical differentiation. L\'opez-Sepulcre et al. (2017) proposed  that the COM-rich protostar is either more massive and/or subject to a higher accretion rate, resulting into a lower envelope mass. Watanabe et al. (2017) suggest that the less massive protostar is related to a younger evolutionary stage in which the hot corino (hot core) is not yet developed so that its dimensions are still very small.
In Cep\,E-mm, the presence of high-velocity SiO jets shows evidence of active mass ejection around both protostars.
The short dynamical timescales (500--$1000\yr$) also indicate that these ejections began recently, so that both sources are still in an early evolutionary stage.
Incidentally, Lykke et al. (2015) found an apparent correlation between the source luminosities and the relative abundance of complex organic molecules in a sample of sources including high-mass protostars. The authors have suggested that this could be the result of the time scale and the temperature a source goes by during its evolution. The sample of sources with evidence for chemical differentiation should be increased  in order to confirm this observational trend.

\subsection{Comparison with hot cores and corinos}

\begin{table}
	\caption{\label{tab:sources} Physical properties of the source sample: distance $d$, bolometric luminosity $L$, envelope mass $M$ and the hot corino (hot core) region size.}
	\begin{tabular}{lcccc}
	
\hline\hline
Source & \textit{d} & \textit{L} & \textit{M} & \textit{size} \\
 & (pc) & ($\lsol$) & ($\msol$) & (au) \\
\hline
	IRAS4A\tablefootmark{a} & 235 & 9.1 & 5.6 & 48 \\
	IRAS16293\tablefootmark{b} & 120 & 22 & 2 & 120 \\
	IRAS2A\tablefootmark{a} & 235 & 36 & 5.1 & 80 \\
	Cep\,E-A & 730 & 100 & 35 & 510 \\
	NGC\,7129\tablefootmark{c} & 1250 & 500 & 50 & 760 \\
	G29.96\tablefootmark{d} & 6000 & $9\times10^4$ & 2500 & 3300 \\
	Orion KL\tablefootmark{d} & 450 & $10^5$ & 140 & 560 \\
	SgrB2(N)\tablefootmark{e} & 7900 & $8.6\times10^5$ & 3 -- $10\times 10^4$ & $2\times10^4$ \\
\hline
	\end{tabular}\\
	\tablefoottext{a}{Taquet et al. 2015} \\
	\tablefoottext{b}{Jaber et al. 2014} \\
	\tablefoottext{c}{Fuente et al. 2005} \\
    \tablefoottext{d}{Beuther et al. 2009}\\
%	\tablefoottext{e}{Pagani et al. 2017} \\
    \tablefoottext{e}{Belloche et al. 2013}
\end{table}

\begin{table*}
\caption[]{\label{tab:rotd-comp} Molecular abundances with respect to \htwo\ measured towards Cep-E\,A and a few protostars of low- (L-M), intermediate- (I-M) and high-mass (H-M) reported from the literature. Sources are ordered from left to right by increasing luminosity.}
\begin{tabular}{@{\extracolsep{4pt}}lcccccccc@{}} \hline\hline	
Molecule		& \multicolumn{8}{c}{$X$} \\
\cline{2-9}
				& \multicolumn{3}{c}{L-M} & \multicolumn{2}{c}{I-M}	& \multicolumn{3}{c}{H-M} \\
\cline{2-4} \cline{5-6} \cline{7-9}
				 &  IRAS4A2\tablefootmark{a} & IRAS16293B\tablefootmark{b}  &  IRAS2A\tablefootmark{a}  &  Cep\,E-A  &  NGC 7129\tablefootmark{c}  &  G29.96\tablefootmark{d} &  Orion KL\tablefootmark{e} & SgrB2(N1)\tablefootmark{f}  \\
 \hline
CH$_3$OH		 &  4.3 (-7)    &  1.7(-6)& 1.0(-6)	 &  1.2(-6)*    &  1.0(-6)*  &  6.7(-8) &  9.3(-8)  &1.4(-6)	\\
H$_2$CO			 &  8.2 (-9)    &  1.6(-7)& 8.0(-8)	 & 8.8(-8)*     &  2.0(-8)*  &  --	   &  --       & 3.9(-8) \\
H$_2$CCO		 &  9.2 (-10)   &  --     & 1.4(-9)	 &  8.0(-9)     &  --	    &  --	   &  9.3(-12) & 4.6(-8) \\
HCOOH	     &  (0.6--2.9)(-9)  &  --     &   --	 &  8.3(-10)    &  3.0(-8)*  &   --	   &  1.3(-10) & 1.2(-9) \\
HCOOCH$_3$		 &  1.1(-8)     & 3.3(-8) & 1.6(-8)	 &  2.6(-8)     &  2.0(-8)   &  1.3(-8) &  4.4(-9)  & 3.4(-8) \\
CH$_3$CHO	     &(1.1--7.4)(-9)& 5.8(-9) &  --		 &  1.1(-8)	    &  2.0(-9)   &  --      &  1.3(-11) & 1.1(-8) \\
CH$_3$OCH$_3$	 &  1.0(-8)	    & --      & 1.0(-8)  &  3.6(-8)     &  1.0(-8)   &  3.3(-8) &  2.6(-9)  & 1.5(-7) \\
CH$_3$COCH$_3$	 &  --		    & 1.4(-9) &    --	 &(2.5--8.3)(-9)&  5.0(-10)  &  --	   &  --	   & 1.2(-8) \\
HNCO			 &(0.3--1.4)(-9)& 2.5(-9) &    --	 &  1.6(-9)     &  8.0(-10)  &  --	   &  1.9(-9)  & 1.0(-7) \\
NH$_2$CHO		 &(1.2-6.7)(-10)& 8.3(-10)&  2.4(-9) &  4.5(-10)    &  --	    &  --	   &  1.9(-10) & 1.0(-7) \\
CH$_3$CN		 &(0.3--1.4)(-9)& 3.3(-9) &  4.0(-9) &  3.5(-9)     &  6.0(-9)*  &  1.7(-9) &  8.5(-10) & 1.5(-7) \\
C$_2$H$_5$CN	&(1.7--5.6)(-10)& 3.0(-10)&  3.0(-10)&(0.45--1.1)(-9)&  1.4(-9)  &  1.7(-9) &  1.3(-9) & 1.4(-7) \\
	\hline
	\end{tabular}\\
* Obtained from the $^{13}$C isotopologue.\\
	\tablefoottext{a}{L\'opez-Sepulcre et al. 2017; Taquet et al. 2015} \\
	\tablefoottext{b}{J{\o}rgensen et al. 2016; Coutens et al. 2016; Lykke et al. 2017; Calcutt et al. 2018} \\
	\tablefoottext{c}{Fuente et al. 2014} \\
    \tablefoottext{d}{Beuther et al. 2009}\\
	\tablefoottext{e}{Pagani et al. 2017, Beuther et al. 2009} \\
    \tablefoottext{f}{Belloche et al. 2013}\\
\end{table*}

\begin{table*}
\caption[]{\label{tab:rotd-comp-meth} COMs abundances with respect to methyl formate. Sources are ordered from left to right by increasing luminosity. }
\begin{tabular}{@{\extracolsep{4pt}}lcccccccc@{}}	\hline\hline	
Molecule		& \multicolumn{8}{c}{$X_{\mathrm{methylformate}}$} \\
	\cline{2-9}
				& 	\multicolumn{3}{c}{L-M} & \multicolumn{2}{c}{I-M}	& \multicolumn{3}{c}{H-M} \\
\cline{2-4} \cline{5-6} \cline{7-9}
			  &  IRAS4A2\tablefootmark{a}  &  IRAS16293B\tablefootmark{b} & IRAS2A\tablefootmark{a}  & Cep\,E-A  &  NGC 7129\tablefootmark{c} &  G29.96\tablefootmark{d} &  Orion KL\tablefootmark{e} & SgrB2(N1)\tablefootmark{f} \\ \hline
CH$_3$OH      & 39       &  51        & 63      & 46        & 50        & 5.2     & 21        & 41 \\
H$_2$CO       & 0.7      &  5         & 5.7     & 3.4       & 1.0       & --      & --        & 1.1 \\
H$_2$CCO      & 0.084    &  --        & 8.8(-2) & 3.1(-1)   & --        & --      & 2.1(-3)   & 1.4 \\
HCOOH         &0.13--0.24&  --        & --      & 3.2(-2)   & 1.5       & --      & 3.0(-2)   & 3.5(-2) \\
CH$_3$CHO     &0.29--0.54&  1.8(-1)   &  --     & 4.2(-1)   & 1.0(-1)   & --      & 3.0(-3)   & 3.2(-1) \\
CH$_3$OCH$_3$ & 0.9      &  --        & 6.3(-1) & 1.4       & 5.0(-1)   & 2.5     & 5.9(-1)   & 4.4 \\
CH$_3$COCH$_3$& --       &  4.0(-2)   & --      & (1.0--3.2)(-1)& 2.5(-2)& --     & --        & 3.5(-1) \\
HNCO         &0.066--0.11&  7.5(-2)   &   --    & 6.1(-2)   & 4.0(-2)   & --      & 4.3(-1)   & 2.9 \\
NH$_2$CHO     &0.028--0.057&  2.5(-2)   & 1.5(-1) & 1.7(-2)   & --        & --      & 4.3(-2)   & 2.9 \\
CH$_3$CN      &0.54--1.4&  1.0(-1)   & 2.5(-1) & 1.3(-1)   & 3.0(-1)   & 1.3(-1) & 1.9(-1)   & 4.4 \\
C$_2$H$_5$CN  & 4.1(-2)  &  1.0(-2)   & 1.9(-2) & (1.7--4.2)(-2)& 7.0(-2)& 1.3(-1)& 3.0(-1)   & 4.1 \\
 \hline
\end{tabular}\\
* Obtained from the $^{13}$C isotopologue.\\
	\tablefoottext{a}{L\'opez-Sepulcre et al. 2017; Taquet et al. 2015} \\
	\tablefoottext{b}{J{\o}rgensen et al. 2016; Coutens et al. 2016; Lykke et al. 2017; Calcutt et al. 2018} \\
	\tablefoottext{c}{Fuente et al. 2014} \\
    \tablefoottext{d}{Beuther et al. 2009}\\
	\tablefoottext{e}{Pagani et al. 2017, Beuther et al. 2009} \\
    \tablefoottext{f}{Belloche et al. 2013}\\
\end{table*}

\begin{figure*}
	\includegraphics[clip, trim=1.3cm 9.cm 1.2cm 2cm, width=2\columnwidth]{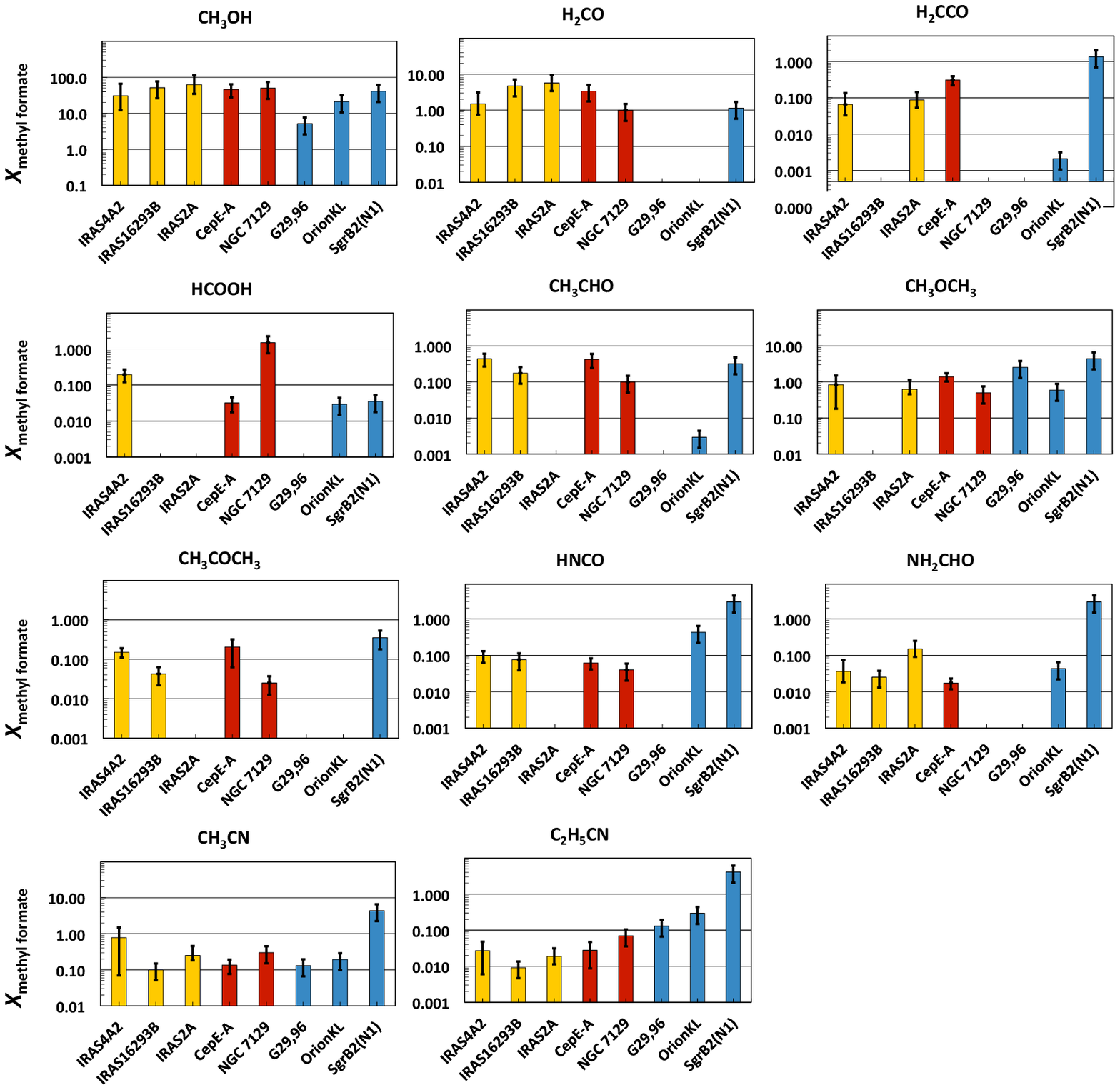}
	\caption{\label{fig:abundances_methanol} Molecular abundances with respect to methyl formate towards low-mass hot corinos (yellow): IRAS2A (Taquet et al. 2015) and IRAS4A2 in NGC1333 (Taquet et al. 2015; L\'opez-Sepulcre et al. 2017 et al.), IRAS16293B (J{\o}rgensen et al. 2016, Coutens et al. 2016, Lykke et al. 2017, Calcutt et al. 2018); the intermediate-mass protostars (red): Cep\,E-A (this work) and NGC~7129 (Fuente et al. 2014); the high-mass hot cores (blue): G29.96 (Beuther et al. 2009), Orion KL  (Pagani et al. 2017, Beuther et al. 2009) and SgrB2(N1) (Belloche et al. 2013). Sources are ordered from left to right by increasing luminosity.}
\end{figure*}

In this section, we compare the molecular abundances of COMs and chemically related species in Cep\,E-A with those obtained towards hot cores and corinos, relatively to $\htwo$ and methyl formate.
The latter is commonly observed in star-forming regions and it does not suffer the drawbacks that affect methanol. First, methanol line emission of low energy levels ($E_{up} < 100\K$) is often strongly contaminated by the outflow and the envelope, as shown in our single-dish analysis of CH$_3$OH emission (Sect.~4.2). Second, in the case of interferometric observations, the amount of filtered flux can vary significantly, hence introducing biases in the derivation of molecular abundances. Finally, methyl formate likely less suffers from opacity issues in comparison to  methanol.

In order to minimize any possible bias in source comparison, we have  selected targets in which the emission of hot cores and corinos could be isolated from the other components contributing to the emission (outflow, envelope). Also, the selected sources  have been investigated in a systematic manner so that a large dataset is available: the I-M protostar NGC 7129 (Fuente et al. 2014), the L-M protostars IRAS2A (Taquet et al. 2015) and IRAS4A2 (L\'opez-Sepulcre et al. 2017; Taquet et al. 2015) in NGC1333  and IRAS16293B (J{\o}rgensen et al. 2016; Coutens et al. 2016; Lykke et al 2017; Persson et al. 2018; Calcutt et al. 2018), the hot cores  G29.96 (Beuther et al. 2009), Orion KL (Beuther et al. 2009, Pagani et al. 2017) and  SgrB2(N1) (Belloche et al. 2013). For all the sources but SgrB2(N1), we made use of interferometric observations.  The molecular content of SgrB2(N1) was obtained by Belloche et al. (2013) from a careful multi-component analysis of a line survey with the IRAM 30m telescope, which allowed them to disentangle the contributions of the two hot cores N1 and N2 of SgrB2(N). In the case of IRAS4A2, the molecular abundances of CH$_3$OH, H$_2$CO and H$_2$CCO were taken from a previous study with the Plateau de Bure Interferometer by Taquet et al. (2015), as those species were not covered by the observations of Lopez-Sepulcre et al. (2017). The angular resolution of their observations ($2\arcsec$) did not allow  Taquet et al. (2015)  to disentangle the contributions of protostars A1 and A2. They estimated the  total gas  column density from thermal dust continuum emission, which is dominated by the protostar A1, whereas the molecular emission arises mainly from the protostar A2. Hence, the abundances of CH$_3$OH, H$_2$CO and H$_2$CCO relatively to \htwo\, as determined by Taquet et al. (2015), are underestimated. On the contrary, the  molecular abundances relatively to methyl formate are not affected by this bias. The physical properties of the parental cores  (distance, luminosity, size) are summarized in Table~\ref{tab:sources}.

The molecular abundances measured  towards Cep\,E-A are summarized in Tables~\ref{tab:rotd-comp} and \ref{tab:rotd-comp-meth}.
In order to facilitate the readibility of Tables 8--9, we report only the molecular abundances.  Abundance uncertainties are indicated  by a black thick line in Fig.~\ref{fig:abundances_methanol}, which  displays a graphical representation of the abundances of COMs with respect to methyl formate. We note that Fuente et al. (2014), Beuther et al. (2009) and Belloche et al. (2013) did not provide any uncertainty  for  molecular abundances  in NGC 7129, Orion KL, G29.96 and SgrB2(N1). Beuther et al. (2009) pointed out that the uncertainties on the source temperature and  sizes imply that the derived column densities must be taken with caution. Therefore, we have adopted a typical relative uncertainty of $\sim 35\%$ for all column density determinations in these sources, similar to our own determinations in CepE-A, but we warn that they could be larger for the high-mass sources.

Overall, the molecular abundances derived in Cep\,E-A agree within a factor of 4 at most with those measured  towards the I-M protostar NGC7129 by Fuente et al. (2005,2014).
Most of the  O- and N-bearing species of our study (CH$_3$CN, C$_2$H$_5$CN and HNCO) have very similar abundances.
The only exceptions are CH$_3$COCH$_3$ and HCOOH. The CH$_3$COCH$_3$ abundance is higher towards Cep\,E-A  by one order of magnitude. We note that this abundance determination also  suffers  large uncertainties  (see Sect.~4.4.2). On the contrary, HCOOH appears more abundant towards NGC~7129 by a factor 30.

Molecular abundances of the L-M  hot corinos IRAS2A and IRAS4A2 are also similar to those of Cep\,E-A.
These conclusions are unchanged when comparing molecular abundances relatively to  methyl formate, as illustrated by the cases of H$_2$CO, HCOOCH$_3$ or CH$_3$OCH$_3$ (Fig.~10).  Our results also agree with the tight correlation between CH$_3$OCH$_3$ and HCOOCH$_3$ previously reported by Jaber et al. (2014).
The molecular abundances of O-bearing species are $\approx 1\%$ of the  CH$_3$OH abundance, whereas those of N-bearing species are lower by one order of magnitude, $\sim 0.1\%$.

The molecular abundances of O-bearing species appear rather similar towards L-M and I-M sources, independently of the luminosity. Towards H-M sources, a large scatter is observed in the relative abundances of HCOOH and CH$_3$CHO.
Comparison of molecular abundances relative to methyl formate does not reveal any specific trend as a function of the source luminosity for O-bearing species, from L-M to H-M.
On the contrary, the abundances of N-bearing species display marked differences. As can be seen in Table 8,  the abundances in G29.96, Orion KL, and  Cep\,E-A are rather similar, whereas SgrB2(N) display  higher abundances. Comparison of the molecular abundances relative to methyl formate reveals a trend,  which is best illustrated by C$_2$H$_5$CN in Fig.~10, namely the relative molecular abundance increases with the source luminosity, from L-M to H-M. The same trend seems to be present in HNCO.

In summary, we find molecular abundances similar between Cep\,E-A , L-M and I-M hot corinos. The relative composition of O-bearing species with respect to methyl formate seems relatively independent of the source luminosity, contrary to C$_2$H$_5$CN, and perhaps other N-bearing species, which increases  as a function of source luminosity.

\section{Conclusions}

We have performed an unbiased spectral survey from 3 to 0.9~mm with the IRAM 30m telescope. It was complemented with interferometric observations in the 3 and 1.3~mm bands with NOEMA at an angular resolution of 2.4\arcsec\ and 1.4\arcsec, respectively. We report bright emission of COMs and chemically related species towards the isolated intermediate-mass protostellar core Cep\,E-mm. Our main conclusions are:

\begin{itemize}

\item Cep\,E-mm appears to be a protostellar binary system with components Cepe\,E-A and Cepe\,E-B separated by $\simeq$ 1.7\arcsec (1250~au). Cep\,E-A dominates the bulk of the continuum emission and powers the long studied high velocity jet associated with HH377. Cep\,E-B powers another high-velocity jet which propagates in a direction almost perpendicular to the Cep\,E-A jet.

\item We found evidence for a hot corino in a region of $\simeq 0.7\arcsec$ around Cep\,E-A. Cep\,E-B seems devoided of molecular emission at the sensitivity of our observations.

\item We successfully identified three components in the molecular spectral signatures : a) an extremely broad line (eBL) component associated with the outflowing gas; b) a narrow line (NL) component associated with the cold outer envelope of Cep\,E-mm; c) a broad line (BL) component which traces the signature of the hot corino.

\item Methanol emission is dominated by the outflowing gas in transitions up to $E_{up} < 100\K$. At higher $E_{up}$, the hot corino dominates over the outflow.

\item Overall, COMs molecular abundances in the Cep\,E-A hot corino are similar to those measured towards other low- and intermediate-mass protostars. N-bearing species are one order of magnitude less abundant that O-bearing species.

\item Relatively to methyl formate, molecular abundances of O-bearing species are rather similar between protostars, independently of the source luminosity.  On the contrary, a good correlation is observed between the relative C$_2$H$_5$CN abundance and the source luminosity.
\end{itemize}

\begin{acknowledgements}
J.O.Z, B.L. and C.F. thank Dr. Laurent Pagani for communication and discussions on Orion-KL.
Based on observations carried out under project number W14AF with the IRAM NOEMA
Interferometer. IRAM is supported by INSU/CNRS (France), MPG (Germany) and IGN (Spain). This work  was supported by a grant from LabeX Osug@2020 (Investissements d'avenir - ANR10LABX56).
C. Favre acknowledge the financial support for this work provided by the French space agency CNES along with the support from the Italian Ministry of Education, Universities and Research, project SIR (RBSI14ZRHR).
\end{acknowledgements}

\nocite{Sargent1977}
\nocite{Wouterloot1986}
\nocite{Langer1992}
\nocite{Palla1993}
\nocite{Lefloch1996}
\nocite{Lefloch1998}
\nocite{Pickett1998}
\nocite{Hogerheidje1999}
\nocite{Ayala2000}
\nocite{Moro-Martin2001}
\nocite{Schreyer2002}
\nocite{Cazaux2003}
%\nocite{Bottinelli2004}
\nocite{Fuente2005}
\nocite{Muller2005}
\nocite{Beuther2007}
\nocite{Neri2007}
\nocite{Fuente2007}
\nocite{Ceccarelli2007}
\nocite{Sakai2008}
\nocite{Shimajiri2008}
\nocite{Beuther2009}
\nocite{Ceccarelli2010}
\nocite{Alonso-albi2010}
\nocite{Pilbratt2010}
\nocite{Kama2010}
\nocite{Ceccarelli2010}
\nocite{Maret2011}
\nocite{Caselli2012}
\nocite{Lopez-Sepulcre2013}
\nocite{Fuente2014}
\nocite{Jaber2014}
\nocite{Lefloch2015}
\nocite{Santangelo2015}
\nocite{Taquet2015}
\nocite{Lopez-Sepulcre2015}
\nocite{Jorgensen2016}
\nocite{Lefloch2017}
\nocite{Watanabe2017}
\nocite{Ceccarelli2017}
\nocite{Lopez-Sepulcre2017}
\nocite{Lefloch2011}
\nocite{Leurini2016}
\nocite{Codella2016}
\nocite{Oya2017}
\nocite{Imai2016}
\nocite{Goldsmith1999}
\nocite{Sanchez-Monge2014}
\nocite{Lefloch2018}
\nocite{Bianchi2017}
\nocite{Oberg2011}
\nocite{Ossenkopf1994}
\nocite{Belloche2008}
\nocite{Guelin2008}
\nocite{Favre2011}
\nocite{Friedel2012}
\nocite{Peng2013}
\nocite{Brouillet2013}
\nocite{Crockett2015}
\nocite{Favre2017}
\nocite{Pagani2017}
\nocite{Lykke2015}
\nocite{Coutens2016}
\nocite{Lykke2017}
\nocite{Persson2018}
\nocite{Calcutt2018}
\nocite{Belloche2013}
\nocite{Qin2015}
\nocite{Crockett2014}
\nocite{Nishimura2017}
\nocite{Bottinelli2004IRAS16293}
\nocite{Bottinelli2004NGC1333}
\nocite{Jorgensen2011}

\bibliographystyle{bibtex/aa}
\bibliography{bibtex/mybib}

\begin{thebibliography}{68}
\expandafter\ifx\csname natexlab\endcsname\relax\def\natexlab#1{#1}\fi

\bibitem[{{Alonso-Albi} {et~al.}(2010){Alonso-Albi}, {Fuente}, {Crimier},
  {Caselli}, {Ceccarelli}, {Johnstone}, {Planesas}, {Rizzo}, {Wyrowski},
  {Tafalla}, {Lefloch}, {Maret}, \& {Dominik}}]{Alonso-albi2010}
{Alonso-Albi}, T., {Fuente}, A., {Crimier}, N., {et~al.} 2010, \aap, 518, A52

\bibitem[{{Ayala} {et~al.}(2000){Ayala}, {Noriega-Crespo}, {Garnavich},
  {Curiel}, {Raga}, {B{\"o}hm}, \& {Raymond}}]{Ayala2000}
{Ayala}, S., {Noriega-Crespo}, A., {Garnavich}, P.~M., {et~al.} 2000, \aj, 120,
  909

\bibitem[{{Belloche} {et~al.}(2008){Belloche}, {Menten}, {Comito},
  {M{\"u}ller}, {Schilke}, {Ott}, {Thorwirth}, \& {Hieret}}]{Belloche2008}
{Belloche}, A., {Menten}, K.~M., {Comito}, C., {et~al.} 2008, \aap, 482, 179

\bibitem[{{Belloche} {et~al.}(2013){Belloche}, {M{\"u}ller}, {Menten},
  {Schilke}, \& {Comito}}]{Belloche2013}
{Belloche}, A., {M{\"u}ller}, H.~S.~P., {Menten}, K.~M., {Schilke}, P., \&
  {Comito}, C. 2013, \aap, 559, A47

\bibitem[{{Beuther} {et~al.}(2009){Beuther}, {Zhang}, {Bergin}, \&
  {Sridharan}}]{Beuther2009}
{Beuther}, H., {Zhang}, Q., {Bergin}, E.~A., \& {Sridharan}, T.~K. 2009, \aj,
  137, 406

\bibitem[{{Beuther} {et~al.}(2007){Beuther}, {Zhang}, {Bergin}, {Sridharan},
  {Hunter}, \& {Leurini}}]{Beuther2007}
{Beuther}, H., {Zhang}, Q., {Bergin}, E.~A., {et~al.} 2007, \aap, 468, 1045

\bibitem[{{Bianchi} {et~al.}(2017){Bianchi}, {Codella}, {Ceccarelli},
  {Fontani}, {Testi}, {Bachiller}, {Lefloch}, {Podio}, \&
  {Taquet}}]{Bianchi2017}
{Bianchi}, E., {Codella}, C., {Ceccarelli}, C., {et~al.} 2017, \mnras, 467,
  3011

\bibitem[{{Brouillet} {et~al.}(2013){Brouillet}, {Despois}, {Baudry}, {Peng},
  {Favre}, {Wootten}, {Remijan}, {Wilson}, {Combes}, \&
  {Wlodarczak}}]{Brouillet2013}
{Brouillet}, N., {Despois}, D., {Baudry}, A., {et~al.} 2013, \aap, 550, 46

\bibitem[{{Calcutt} {et~al.}(2018){Calcutt}, {J{\o}rgensen}, {M{\"u}ller},
  {Kristensen}, {Coutens}, {Bourke}, {Garrod}, {Persson}, {van der Wiel}, {van
  Dishoeck}, \& {Wampfler}}]{Calcutt2018}
{Calcutt}, H., {J{\o}rgensen}, J.~K., {M{\"u}ller}, H.~S.~P., {et~al.} 2018,
  ArXiv e-prints, arXiv:1804.09210

\bibitem[{{Caselli} \& {Ceccarelli}(2012)}]{Caselli2012}
{Caselli}, P. \& {Ceccarelli}, C. 2012, \aapr, 20, 56

\bibitem[{{Cazaux} {et~al.}(2003){Cazaux}, {Tielens}, {Ceccarelli}, {Castets},
  {Wakelam}, {Caux}, {Parise}, \& {Teyssier}}]{Cazaux2003}
{Cazaux}, S., {Tielens}, A.~G.~G.~M., {Ceccarelli}, C., {et~al.} 2003, \apjl,
  593, L51

\bibitem[{{Ceccarelli} {et~al.}(2010){Ceccarelli}, {Bacmann}, {Boogert},
  {Caux}, {Dominik}, {Lefloch}, {Lis}, {Schilke}, {van der Tak}, {Caselli},
  {Cernicharo}, {Codella}, {Comito}, {Fuente}, {Baudry}, {Bell}, {Benedettini},
  {Bergin}, {Blake}, {Bottinelli}, {Cabrit}, {Castets}, {Coutens}, {Crimier},
  {Demyk}, {Encrenaz}, {Falgarone}, {Gerin}, {Goldsmith}, {Helmich},
  {Hennebelle}, {Henning}, {Herbst}, {Hily-Blant}, {Jacq}, {Kahane}, {Kama},
  {Klotz}, {Langer}, {Lord}, {Lorenzani}, {Maret}, {Melnick}, {Neufeld},
  {Nisini}, {Pacheco}, {Pagani}, {Parise}, {Pearson}, {Phillips}, {Salez},
  {Saraceno}, {Schuster}, {Tielens}, {van der Wiel}, {Vastel}, {Viti},
  {Wakelam}, {Walters}, {Wyrowski}, {Yorke}, {Liseau}, {Olberg}, {Szczerba},
  {Benz}, \& {Melchior}}]{Ceccarelli2010}
{Ceccarelli}, C., {Bacmann}, A., {Boogert}, A., {et~al.} 2010, \aap, 521, L22

\bibitem[{{Ceccarelli} {et~al.}(2017){Ceccarelli}, {Caselli}, {Fontani},
  {Neri}, {L{\'o}pez-Sepulcre}, {Codella}, {Feng}, {Jim{\'e}nez-Serra},
  {Lefloch}, {Pineda}, {Vastel}, {Alves}, {Bachiller}, {Balucani}, {Bianchi},
  {Bizzocchi}, {Bottinelli}, {Caux}, {Chac{\'o}n-Tanarro}, {Choudhury},
  {Coutens}, {Dulieu}, {Favre}, {Hily-Blant}, {Holdship}, {Kahane}, {Jaber
  Al-Edhari}, {Laas}, {Ospina-Zamudio}, {Oya}, {Podio}, {Pon}, {Punanova},
  {Quenard}, {Rimola}, {Sakai}, {Sims}, {Spezzano}, {Taquet}, {Testi},
  {Theul{\'e}}, {Ugliengo}, {Vasyunin}, {Viti}, {Wiesenfeld}, \&
  {Yamamoto}}]{Ceccarelli2017}
{Ceccarelli}, C., {Caselli}, P., {Fontani}, F., {et~al.} 2017, \apj, 850, 176

\bibitem[{{Ceccarelli} {et~al.}(2007){Ceccarelli}, {Caselli}, {Herbst},
  {Tielens}, \& {Caux}}]{Ceccarelli2007}
{Ceccarelli}, C., {Caselli}, P., {Herbst}, E., {Tielens}, A.~G.~G.~M., \&
  {Caux}, E. 2007, Protostars and Planets V, 47

\bibitem[{{Codella} {et~al.}(2016){Codella}, {Ceccarelli}, {Cabrit}, {Gueth},
  {Podio}, {Bachiller}, {Fontani}, {Gusdorf}, {Lefloch}, {Leurini}, \&
  {Tafalla}}]{Codella2016}
{Codella}, C., {Ceccarelli}, C., {Cabrit}, S., {et~al.} 2016, \aap, 586, L3

\bibitem[{{Coutens} {et~al.}(2016){Coutens}, {J{\o}rgensen}, {van der Wiel},
  {M{\"u}ller}, {Lykke}, {Bjerkeli}, {Bourke}, {Calcutt}, {Drozdovskaya},
  {Favre}, {Fayolle}, {Garrod}, {Jacobsen}, {Ligterink}, {{\"O}berg},
  {Persson}, {van Dishoeck}, \& {Wampfler}}]{Coutens2016}
{Coutens}, A., {J{\o}rgensen}, J.~K., {van der Wiel}, M.~H.~D., {et~al.} 2016,
  \aap, 590, L6

\bibitem[{{Crockett} {et~al.}(2015){Crockett}, {Bergin}, {Neill}, {Favre},
  {Blake}, {Herbst}, {Anderson}, \& {Hassel}}]{Crockett2015}
{Crockett}, N.~R., {Bergin}, E.~A., {Neill}, J.~L., {et~al.} 2015, \apj, 806,
  239

\bibitem[{{Crockett} {et~al.}(2014){Crockett}, {Bergin}, {Neill}, {Favre},
  {Schilke}, {Lis}, {Bell}, {Blake}, {Cernicharo}, {Emprechtinger},
  {Esplugues}, {Gupta}, {Kleshcheva}, {Lord}, {Marcelino}, {McGuire},
  {Pearson}, {Phillips}, {Plume}, {van der Tak}, {Tercero}, \&
  {Yu}}]{Crockett2014}
{Crockett}, N.~R., {Bergin}, E.~A., {Neill}, J.~L., {et~al.} 2014, \apj, 787,
  112

\bibitem[{{Favre} {et~al.}(2011){Favre}, {Despois}, {Brouillet}, {Baudry},
  {Combes}, {Gu{\'e}lin}, {Wootten}, \& {Wlodarczak}}]{Favre2011}
{Favre}, C., {Despois}, D., {Brouillet}, N., {et~al.} 2011, \aap, 532, A32

\bibitem[{{Favre} {et~al.}(2017){Favre}, {Pagani}, {Goldsmith}, {Bergin},
  {Carvajal}, {Kleiner}, {Melnick}, \& {Snell}}]{Favre2017}
{Favre}, C., {Pagani}, L., {Goldsmith}, P.~F., {et~al.} 2017, \aap, 604, L2

\bibitem[{{Friedel} \& {Widicus Weaver}(2012)}]{Friedel2012}
{Friedel}, D.~N. \& {Widicus Weaver}, S.~L. 2012, The Astrophysical Journal
  Supplement Series, 201, 17

\bibitem[{{Fuente} {et~al.}(2007){Fuente}, {Ceccarelli}, {Neri}, {Alonso-Albi},
  {Caselli}, {Johnstone}, {van Dishoeck}, \& {Wyrowski}}]{Fuente2007}
{Fuente}, A., {Ceccarelli}, C., {Neri}, R., {et~al.} 2007, \aap, 468, L37

\bibitem[{{Fuente} {et~al.}(2014){Fuente}, {Cernicharo}, {Caselli}, {McCoey},
  {Johnstone}, {Fich}, {van Kempen}, {Palau}, {Y{\i}ld{\i}z}, {Tercero}, \&
  {L{\'o}pez}}]{Fuente2014}
{Fuente}, A., {Cernicharo}, J., {Caselli}, P., {et~al.} 2014, \aap, 568, A65

\bibitem[{{Fuente} {et~al.}(2005){Fuente}, {Neri}, \& {Caselli}}]{Fuente2005}
{Fuente}, A., {Neri}, R., \& {Caselli}, P. 2005, \aap, 444, 481

\bibitem[{{Goldsmith} \& {Langer}(1999)}]{Goldsmith1999}
{Goldsmith}, P.~F. \& {Langer}, W.~D. 1999, \apj, 517, 209

\bibitem[{{Gu{\'e}lin} {et~al.}(2008){Gu{\'e}lin}, {Brouillet}, {Cernicharo},
  {Combes}, \& {Wootten}}]{Guelin2008}
{Gu{\'e}lin}, M., {Brouillet}, N., {Cernicharo}, J., {Combes}, F., \&
  {Wootten}, A. 2008, \apss, 313, 45

\bibitem[{{Hogerheijde} {et~al.}(1999){Hogerheijde}, {van Dishoeck},
  {Salverda}, \& {Blake}}]{Hogerheidje1999}
{Hogerheijde}, M.~R., {van Dishoeck}, E.~F., {Salverda}, J.~M., \& {Blake},
  G.~A. 1999, \apj, 513, 350

\bibitem[{{Imai} {et~al.}(2016){Imai}, {Sakai}, {Oya}, {L{\'o}pez- Sepulcre},
  {Watanabe}, {Ceccarelli}, {Lefloch}, {Caux}, {Vastel}, {Kahane}, {Sakai},
  {Hirota}, {Aikawa}, \& {Yamamoto}}]{Imai2016}
{Imai}, M., {Sakai}, N., {Oya}, Y., {et~al.} 2016, \apj, 830, L37

\bibitem[{{Jaber} {et~al.}(2014){Jaber}, {Ceccarelli}, {Kahane}, \&
  {Caux}}]{Jaber2014}
{Jaber}, A.~A., {Ceccarelli}, C., {Kahane}, C., \& {Caux}, E. 2014, \apj, 791,
  29

\bibitem[{{J{\o}rgensen} {et~al.}(2016){J{\o}rgensen}, {van der Wiel},
  {Coutens}, {Lykke}, {M{\"u}ller}, {van Dishoeck}, {Calcutt}, {Bjerkeli},
  {Bourke}, {Drozdovskaya}, {Favre}, {Fayolle}, {Garrod}, {Jacobsen},
  {{\"O}berg}, {Persson}, \& {Wampfler}}]{Jorgensen2016}
{J{\o}rgensen}, J.~K., {van der Wiel}, M.~H.~D., {Coutens}, A., {et~al.} 2016,
  \aap, 595, A117

\bibitem[{{Kama} {et~al.}(2010){Kama}, {Dominik}, {Maret}, {van der Tak},
  {Caux}, {Ceccarelli}, {Fuente}, {Crimier}, {Lord}, {Bacmann}, {Baudry},
  {Bell}, {Benedettini}, {Bergin}, {Blake}, {Boogert}, {Bottinelli}, {Cabrit},
  {Caselli}, {Castets}, {Cernicharo}, {Codella}, {Comito}, {Coutens}, {Demyk},
  {Encrenaz}, {Falgarone}, {Gerin}, {Goldsmith}, {Helmich}, {Hennebelle},
  {Henning}, {Herbst}, {Hily-Blant}, {Jacq}, {Kahane}, {Klotz}, {Langer},
  {Lefloch}, {Lis}, {Lorenzani}, {Melnick}, {Nisini}, {Pacheco}, {Pagani},
  {Parise}, {Pearson}, {Phillips}, {Salez}, {Saraceno}, {Schilke}, {Schuster},
  {Tielens}, {van der Wiel}, {Vastel}, {Viti}, {Wakelam}, {Walters},
  {Wyrowski}, {Yorke}, {Cais}, {G{\"u}sten}, {Philipp}, {Klein}, \&
  {Helmich}}]{Kama2010}
{Kama}, M., {Dominik}, C., {Maret}, S., {et~al.} 2010, \aap, 521, L39

\bibitem[{{Langer}(1992)}]{Langer1992}
{Langer}, W.~D. 1992, in IAU Symposium, Vol. 150, Astrochemistry of Cosmic
  Phenomena, ed. P.~D. {Singh}, 193

\bibitem[{{Lefloch} {et~al.}(2018){Lefloch}, {Bachiller}, {Ceccarelli},
  {Cernicharo}, {Codella}, {Fuente}, {Kahane}, {L{\'o}pez-Sepulcre}, {Tafalla},
  {Vastel}, {Caux}, {Gonz{\'a}lez-Garc{\'\i}a}, {Bianchi}, {G{\'o}mez-Ruiz},
  {Holdship}, {Mendoza}, {Ospina-Zamudio}, {Podio}, {Qu{\'e}nard}, {Roueff},
  {Sakai}, {Viti}, {Yamamoto}, {Yoshida}, {Favre}, {Monfredini},
  {Quiti{\'a}n-Lara}, {Marcelino}, {Boechat-Roberty}, \&
  {Cabrit}}]{Lefloch2018}
{Lefloch}, B., {Bachiller}, R., {Ceccarelli}, C., {et~al.} 2018, \mnras, 477,
  4792

\bibitem[{{Lefloch} {et~al.}(1998){Lefloch}, {Castets}, {Cernicharo}, \&
  {Loinard}}]{Lefloch1998}
{Lefloch}, B., {Castets}, A., {Cernicharo}, J., \& {Loinard}, L. 1998, \apjl,
  504, L109

\bibitem[{{Lefloch} {et~al.}(2017){Lefloch}, {Ceccarelli}, {Codella}, {Favre},
  {Podio}, {Vastel}, {Viti}, \& {Bachiller}}]{Lefloch2017}
{Lefloch}, B., {Ceccarelli}, C., {Codella}, C., {et~al.} 2017, \mnras, 469, L73

\bibitem[{{Lefloch} {et~al.}(2011){Lefloch}, {Cernicharo}, {Pacheco}, \&
  {Ceccarelli}}]{Lefloch2011}
{Lefloch}, B., {Cernicharo}, J., {Pacheco}, S., \& {Ceccarelli}, C. 2011, \aap,
  527, L3

\bibitem[{{Lefloch} {et~al.}(1996){Lefloch}, {Eisloeffel}, \&
  {Lazareff}}]{Lefloch1996}
{Lefloch}, B., {Eisloeffel}, J., \& {Lazareff}, B. 1996, \aap, 313, L17

\bibitem[{{Lefloch} {et~al.}(2015){Lefloch}, {Gusdorf}, {Codella},
  {Eisl{\"o}ffel}, {Neri}, {G{\'o}mez-Ruiz}, {G{\"u}sten}, {Leurini},
  {Risacher}, \& {Benedettini}}]{Lefloch2015}
{Lefloch}, B., {Gusdorf}, A., {Codella}, C., {et~al.} 2015, \aap, 581, A4

\bibitem[{{Leurini} {et~al.}(2016){Leurini}, {Codella}, {Cabrit}, {Gueth},
  {Giannetti}, {Bacciotti}, {Bachiller}, {Ceccarelli}, {Gusdorf}, {Lefloch},
  {Podio}, \& {Tafalla}}]{Leurini2016}
{Leurini}, S., {Codella}, C., {Cabrit}, S., {et~al.} 2016, \aap, 595, L4

\bibitem[{{L{\'o}pez-Sepulcre} {et~al.}(2015){L{\'o}pez-Sepulcre}, {Jaber},
  {Mendoza}, {Lefloch}, {Ceccarelli}, {Vastel}, {Bachiller}, {Cernicharo},
  {Codella}, {Kahane}, {Kama}, \& {Tafalla}}]{Lopez-Sepulcre2015}
{L{\'o}pez-Sepulcre}, A., {Jaber}, A.~A., {Mendoza}, E., {et~al.} 2015, \mnras,
  449, 2438

\bibitem[{{L{\'o}pez-Sepulcre} {et~al.}(2017){L{\'o}pez-Sepulcre}, {Sakai},
  {Neri}, {Imai}, {Oya}, {Ceccarelli}, {Higuchi}, {Aikawa}, {Bottinelli},
  {Caux}, {Hirota}, {Kahane}, {Lefloch}, {Vastel}, {Watanabe}, \&
  {Yamamoto}}]{Lopez-Sepulcre2017}
{L{\'o}pez-Sepulcre}, A., {Sakai}, N., {Neri}, R., {et~al.} 2017, \aap, 606,
  A121

\bibitem[{{L{\'o}pez-Sepulcre} {et~al.}(2013){L{\'o}pez-Sepulcre}, {Taquet},
  {S{\'a}nchez-Monge}, {Ceccarelli}, {Dominik}, {Kama}, {Caux}, {Fontani},
  {Fuente}, {Ho}, {Neri}, \& {Shimajiri}}]{Lopez-Sepulcre2013}
{L{\'o}pez-Sepulcre}, A., {Taquet}, V., {S{\'a}nchez-Monge}, {\'A}., {et~al.}
  2013, \aap, 556, A62

\bibitem[{{Lykke} {et~al.}(2017){Lykke}, {Coutens}, {J{\o}rgensen}, {van der
  Wiel}, {Garrod}, {M{\"u}ller}, {Bjerkeli}, {Bourke}, {Calcutt},
  {Drozdovskaya}, {Favre}, {Fayolle}, {Jacobsen}, {{\"O}berg}, {Persson}, {van
  Dishoeck}, \& {Wampfler}}]{Lykke2017}
{Lykke}, J.~M., {Coutens}, A., {J{\o}rgensen}, J.~K., {et~al.} 2017, \aap, 597,
  A53

\bibitem[{{Lykke} {et~al.}(2015){Lykke}, {Favre}, {Bergin}, \&
  {J{\o}rgensen}}]{Lykke2015}
{Lykke}, J.~M., {Favre}, C., {Bergin}, E.~A., \& {J{\o}rgensen}, J.~K. 2015,
  \aap, 582, A64

\bibitem[{{Maret} {et~al.}(2011){Maret}, {Hily-Blant}, {Pety}, {Bardeau}, \&
  {Reynier}}]{Maret2011}
{Maret}, S., {Hily-Blant}, P., {Pety}, J., {Bardeau}, S., \& {Reynier}, E.
  2011, \aap, 526, A47

\bibitem[{{Moro-Mart{\'{\i}}n} {et~al.}(2001){Moro-Mart{\'{\i}}n},
  {Noriega-Crespo}, {Molinari}, {Testi}, {Cernicharo}, \&
  {Sargent}}]{Moro-Martin2001}
{Moro-Mart{\'{\i}}n}, A., {Noriega-Crespo}, A., {Molinari}, S., {et~al.} 2001,
  \apj, 555, 146

\bibitem[{{M{\"u}ller} {et~al.}(2005){M{\"u}ller}, {Schl{\"o}der}, {Stutzki},
  \& {Winnewisser}}]{Muller2005}
{M{\"u}ller}, H.~S.~P., {Schl{\"o}der}, F., {Stutzki}, J., \& {Winnewisser}, G.
  2005, Journal of Molecular Structure, 742, 215

\bibitem[{{Neri} {et~al.}(2007){Neri}, {Fuente}, {Ceccarelli}, {Caselli},
  {Johnstone}, {van Dishoeck}, {Wyrowski}, {Tafalla}, {Lefloch}, \&
  {Plume}}]{Neri2007}
{Neri}, R., {Fuente}, A., {Ceccarelli}, C., {et~al.} 2007, \aap, 468, L33

\bibitem[{{Nishimura} {et~al.}(2017){Nishimura}, {Watanabe}, {Harada},
  {Shimonishi}, {Sakai}, {Aikawa}, {Kawamura}, \& {Yamamoto}}]{Nishimura2017}
{Nishimura}, Y., {Watanabe}, Y., {Harada}, N., {et~al.} 2017, \apj, 848, 17

\bibitem[{{{\"O}berg} {et~al.}(2011){{\"O}berg}, {van der Marel}, {Kristensen},
  \& {van Dishoeck}}]{Oberg2011}
{{\"O}berg}, K.~I., {van der Marel}, N., {Kristensen}, L.~E., \& {van
  Dishoeck}, E.~F. 2011, \apj, 740, 14

\bibitem[{{Ossenkopf} \& {Henning}(1994)}]{Ossenkopf1994}
{Ossenkopf}, V. \& {Henning}, T. 1994, \aap, 291, 943

\bibitem[{{Oya} {et~al.}(2017){Oya}, {Sakai}, {Watanabe}, {Higuchi}, {Hirota},
  {L{\'o}pez-Sepulcre}, {Sakai}, {Aikawa}, {Ceccarelli}, {Lefloch}, {Caux},
  {Vastel}, {Kahane}, \& {Yamamoto}}]{Oya2017}
{Oya}, Y., {Sakai}, N., {Watanabe}, Y., {et~al.} 2017, \apj, 837, 174

\bibitem[{{Pagani} {et~al.}(2017){Pagani}, {Favre}, {Goldsmith}, {Bergin},
  {Snell}, \& {Melnick}}]{Pagani2017}
{Pagani}, L., {Favre}, C., {Goldsmith}, P.~F., {et~al.} 2017, \aap, 604, A32

\bibitem[{{Palla} {et~al.}(1993){Palla}, {Cesaroni}, {Brand}, {Caselli},
  {Comoretto}, \& {Felli}}]{Palla1993}
{Palla}, F., {Cesaroni}, R., {Brand}, J., {et~al.} 1993, \aap, 280, 599

\bibitem[{{Peng} {et~al.}(2013){Peng}, {Despois}, {Brouillet}, {Baudry},
  {Favre}, {Remijan}, {Wootten}, {Wilson}, {Combes}, \&
  {Wlodarczak}}]{Peng2013}
{Peng}, T.~C., {Despois}, D., {Brouillet}, N., {et~al.} 2013, \aap, 554, A78

\bibitem[{{Persson} {et~al.}(2018){Persson}, {J{\o}rgensen}, {M{\"u}ller},
  {Coutens}, {van Dishoeck}, {Taquet}, {Calcutt}, {van der Wiel}, {Bourke}, \&
  {Wampfler}}]{Persson2018}
{Persson}, M.~V., {J{\o}rgensen}, J.~K., {M{\"u}ller}, H.~S.~P., {et~al.} 2018,
  \aap, 610, A54

\bibitem[{{Pickett} {et~al.}(1998){Pickett}, {Poynter}, {Cohen}, {Delitsky},
  {Pearson}, \& {M{\"u}ller}}]{Pickett1998}
{Pickett}, H.~M., {Poynter}, R.~L., {Cohen}, E.~A., {et~al.} 1998, \jqsrt, 60,
  883

\bibitem[{{Pilbratt} {et~al.}(2010){Pilbratt}, {Riedinger}, {Passvogel},
  {Crone}, {Doyle}, {Gageur}, {Heras}, {Jewell}, {Metcalfe}, {Ott}, \&
  {Schmidt}}]{Pilbratt2010}
{Pilbratt}, G.~L., {Riedinger}, J.~R., {Passvogel}, T., {et~al.} 2010, \aap,
  518, L1

\bibitem[{{Qin} {et~al.}(2015){Qin}, {Schilke}, {Wu}, {Wu}, {Liu}, {Liu}, \&
  {S{\'a}nchez-Monge}}]{Qin2015}
{Qin}, S.-L., {Schilke}, P., {Wu}, J., {et~al.} 2015, \apj, 803, 39

\bibitem[{{Sakai} {et~al.}(2008){Sakai}, {Sakai}, \& {Yamamoto}}]{Sakai2008}
{Sakai}, N., {Sakai}, T., \& {Yamamoto}, S. 2008, \apss, 313, 153

\bibitem[{{S{\'a}nchez-Monge} {et~al.}(2014){S{\'a}nchez-Monge}, {Beltr{\'a}n},
  {Cesaroni}, {Etoka}, {Galli}, {Kumar}, {Moscadelli}, {Stanke}, {van der Tak},
  {Vig}, {Walmsley}, {Wang}, {Zinnecker}, {Elia}, {Molinari}, \&
  {Schisano}}]{Sanchez-Monge2014}
{S{\'a}nchez-Monge}, {\'A}., {Beltr{\'a}n}, M.~T., {Cesaroni}, R., {et~al.}
  2014, \aap, 569, A11

\bibitem[{{Santangelo} {et~al.}(2015){Santangelo}, {Codella}, {Cabrit},
  {Maury}, {Gueth}, {Maret}, {Lefloch}, {Belloche}, {Andr{\'e}}, {Hennebelle},
  {Anderl}, {Podio}, \& {Testi}}]{Santangelo2015}
{Santangelo}, G., {Codella}, C., {Cabrit}, S., {et~al.} 2015, \aap, 584, A126

\bibitem[{{Sargent}(1977)}]{Sargent1977}
{Sargent}, A.~I. 1977, \apj, 218, 736

\bibitem[{{Schreyer} {et~al.}(2002){Schreyer}, {Henning}, {van der Tak},
  {Boonman}, \& {van Dishoeck}}]{Schreyer2002}
{Schreyer}, K., {Henning}, T., {van der Tak}, F.~F.~S., {Boonman}, A.~M.~S., \&
  {van Dishoeck}, E.~F. 2002, \aap, 394, 561

\bibitem[{{Shimajiri} {et~al.}(2008){Shimajiri}, {Takahashi}, {Takakuwa},
  {Saito}, \& {Kawabe}}]{Shimajiri2008}
{Shimajiri}, Y., {Takahashi}, S., {Takakuwa}, S., {Saito}, M., \& {Kawabe}, R.
  2008, \apj, 683, 255

\bibitem[{{Taquet} {et~al.}(2015){Taquet}, {L{\'o}pez-Sepulcre}, {Ceccarelli},
  {Neri}, {Kahane}, \& {Charnley}}]{Taquet2015}
{Taquet}, V., {L{\'o}pez-Sepulcre}, A., {Ceccarelli}, C., {et~al.} 2015, \apj,
  804, 81

\bibitem[{{Watanabe} {et~al.}(2017){Watanabe}, {Sakai}, {L{\'o}pez-Sepulcre},
  {Sakai}, {Hirota}, {Liu}, {Su}, \& {Yamamoto}}]{Watanabe2017}
{Watanabe}, Y., {Sakai}, N., {L{\'o}pez-Sepulcre}, A., {et~al.} 2017, \apj,
  847, 108

\bibitem[{{Wouterloot} \& {Walmsley}(1986)}]{Wouterloot1986}
{Wouterloot}, J.~G.~A. \& {Walmsley}, C.~M. 1986, \aap, 168, 237

\end{thebibliography}

\begin{appendix}
\section{Spectral properties of detected COMs emission lines.}

\begin{table*}[htbp]
\caption{E-CH3OH emission lines from the hot corino detected with the IRAM~30m telescope. Intensities are expressed in units of $T_A^*$.}
\begin{tabular}{rrrrrrrrr}\hline \hline
Species & Frequency & Quantum numbers & $E_{up}$ & $A_{ul}$ & Flux & V$_{lsr}$ & $\Delta$V & $T_{peak}$ \\
 & (MHz) &  & (K) & (10$^{-5}$ s$^{-1})$ & (mK.km.s$^{-1}$) & (km.s$^{-1}$) & (km.s$^{-1}$) & (mK) \\ \hline
E-CH$_3$OH & 84521.172 & 5$_{-1,0}$ -- 4$_{0,0}$ & 32.5 & 0.2 & 1091 (21) & -12.7 (0.0) & 5.2 (0.1) & 198 (3) \\
 & 108893.945 & 0$_{0,0}$ -- 1$_{-1,0}$ & 5.2 & 1.5 & 160 (18) & -11.9 (0.1) & 4.3 (0.3) & 35 (3) \\
 & 132890.759 & 6$_{-1,0}$ -- 5$_{0,0}$ & 46.4 & 0.8 & 1178 (44) & -12.5 (0.0) & 5.3 (0.1) & 208 (6) \\
 & 145097.435 & 3$_{-1,0}$ -- 2$_{-1,0}$ & 11.6 & 1.1 & 3854 (210) & -12.3 (0.1) & 7.4 (0.4) & 486 (12) \\
 & 145126.386 & 3$_{-2,0}$ -- 2$_{-2,0}$ & 31.9 & 0.7 & 96 (35) & -12.2 (0.5) & 4.5 (0.8) & 20 (6) \\
 & 145131.864 & 3$_{1,0}$ -- 2$_{1,0}$ & 27.1 & 1.1 & 562 (33) & -12.3 (0.1) & 6.1 (0.2) & 87 (4) \\
 & 145766.227 & 16$_{0,0}$ -- 16$_{-1,0}$ & 319.7 & 0.8 & 41 (8) & -10.0 (0.4) & 6.4 (1.0) & 6 (1) \\
 & 150141.672 & 14$_{0,0}$ -- 14$_{-1,0}$ & 248.2 & 1.0 & 38 (10) & -10.7 (0.4) & 4.2 (0.9) & 9 (2) \\
 & 153281.282 & 12$_{0,0}$ -- 12$_{-1,0}$ & 185.9 & 1.3 & 43 (8) & -10.4 (0.2) & 2.9 (0.5) & 14 (2) \\
 & 154425.832 & 11$_{0,0}$ -- 11$_{-1,0}$ & 158.2 & 1.4 & 80 (9) & -10.6 (0.1) & 4.5 (0.4) & 17 (1) \\
 & 155320.895 & 10$_{0,0}$ -- 10$_{-1,0}$ & 132.7 & 1.5 & 83 (12) & -10.7 (0.2) & 5.4 (0.7) & 14 (1) \\
 & 155997.524 & 9$_{0,0}$ -- 9$_{-1,0}$ & 109.6 & 1.7 & 138 (8) & -10.9 (0.1) & 5.2 (0.2) & 25 (1) \\
 & 156488.902 & 8$_{0,0}$ -- 8$_{-1,0}$ & 88.7 & 1.8 & 102 (31) & -10.7 (0.1) & 3.5 (0.6) & 27 (7) \\
 & 156828.517 & 7$_{0,0}$ -- 7$_{-1,0}$ & 70.2 & 1.9 & 139 (9) & -10.9 (0.0) & 3.4 (0.2) & 38 (2) \\
 & 157048.617 & 6$_{0,0}$ -- 6$_{-1,0}$ & 54.0 & 2.0 & 147 (15) & -10.9 (0.1) & 3.1 (0.2) & 44 (4) \\
 & 157178.987 & 5$_{0,0}$ -- 5$_{-1,0}$ & 40.0 & 2.0 & 160 (88) & -11.1 (0.2) & 5.1 (1.7) & 29 (13) \\
 & 157246.062 & 4$_{0,0}$ -- 4$_{-1,0}$ & 28.4 & 2.1 & 181 (12) & -10.9 (0.0) & 2.9 (0.1) & 59 (3) \\
 & 165050.175 & 1$_{1,0}$ -- 1$_{0,0}$ & 15.5 & 2.3 & 533 (24) & -12.0 (0.1) & 6.9 (0.2) & 72 (2) \\
 & 165061.130 & 2$_{1,0}$ -- 2$_{0,0}$ & 20.1 & 2.3 & 219 (34) & -11.5 (0.1) & 3.3 (0.4) & 62 (7) \\
 & 165099.240 & 3$_{1,0}$ -- 3$_{0,0}$ & 27.1 & 2.3 & 269 (63) & -11.5 (0.1) & 4.0 (0.6) & 64 (11) \\
 & 165190.475 & 4$_{1,0}$ -- 4$_{0,0}$ & 36.4 & 2.3 & 201 (45) & -11.3 (0.1) & 3.3 (0.5) & 57 (9) \\
 & 165369.341 & 5$_{1,0}$ -- 5$_{0,0}$ & 48.0 & 2.3 & 251 (37) & -11.2 (0.1) & 3.6 (0.4) & 66 (7) \\
 & 166169.098 & 7$_{1,0}$ -- 7$_{0,0}$ & 78.2 & 2.3 & 190 (48) & -11.3 (0.2) & 6.4 (0.8) & 28 (6) \\
 & 166898.566 & 8$_{1,0}$ -- 8$_{0,0}$ & 96.7 & 2.3 & 202 (12) & -11.0 (0.1) & 4.5 (0.2) & 42 (2) \\
 & 167931.056 & 9$_{1,0}$ -- 9$_{0,0}$ & 117.6 & 2.3 & 150 (13) & -10.4 (0.1) & 4.3 (0.3) & 33 (2) \\
 & 168577.831 & 4$_{1,0}$ -- 3$_{2,0}$ & 36.4 & 0.4 & 117 (47) & -11.6 (0.6) & 8.6 (1.8) & 13 (4) \\
 & 170060.592 & 3$_{2,0}$ -- 2$_{1,0}$ & 28.3 & 2.5 & 1433 (35) & -11.9 (0.0) & 6.9 (0.1) & 195 (4) \\
 & 213427.061 & 1$_{1,0}$ -- 0$_{0,0}$ & 15.5 & 3.4 & 286 (34) & -11.5 (0.3) & 7.0 (0.6) & 38 (3) \\
 & 217886.504 & 20$_{1,0}$ -- 20$_{0,0}$ & 500.5 & 3.4 & 98 (48) & -11.0 (0.9) & 5.9 (2.2) & 16 (5) \\
 & 218440.063 & 4$_{2,0}$ -- 3$_{1,0}$ & 37.6 & 4.7 & 688 (484) & -10.7 (0.6) & 6.1 (1.1) & 106 (72) \\
 & 229758.756 & 8$_{-1,0}$ -- 7$_{0,0}$ & 81.2 & 4.2 & 509 (56) & -11.2 (0.1) & 5.6 (0.4) & 86 (7) \\
 & 232945.797 & 10$_{-3,0}$ -- 11$_{-2,0}$ & 182.5 & 2.1 & 76 (38) & -10.2 (1.2) & 8.0 (3.2) & 9 (3) \\
 & 240241.490 & 5$_{3,0}$ -- 6$_{2,0}$ & 74.6 & 1.4 & 54 (47) & -10.9 (1.0) & 5.1 (2.7) & 10 (7) \\
 & 241179.886 & 5$_{-3,1}$ -- 4$_{-3,1}$ & 349.5 & 3.8 & 67 (28) & -11.0 (0.7) & 5.9 (1.9) & 11 (3) \\
 & 241700.159 & 5$_{0,0}$ -- 4$_{0,0}$ & 40.0 & 6.0 & 828 (51) & -11.2 (0.1) & 6.5 (0.3) & 119 (5) \\
 & 241767.234 & 5$_{-1,0}$ -- 4$_{-1,0}$ & 32.5 & 5.8 & 1677 (98) & -11.2 (0.1) & 6.2 (0.2) & 252 (12) \\
 & 241843.604 & 5$_{3,0}$ -- 4$_{3,0}$ & 74.6 & 3.9 & 229 (13) & -9.5 (0.1) & 6.3 (0.3) & 34 (1) \\
 & 241852.299 & 5$_{-3,0}$ -- 4$_{-3,0}$ & 89.6 & 3.9 & 147 (40) & -12.4 (0.6) & 7.9 (1.7) & 18 (3) \\
 & 241879.025 & 5$_{1,0}$ -- 4$_{1,0}$ & 48.0 & 6.0 & 461 (46) & -11.1 (0.3) & 6.1 (0.5) & 71 (4) \\
 & 241904.147 & 5$_{-2,0}$ -- 4$_{-2,0}$ & 52.8 & 5.1 & 619 (193) & -11.5 (0.2) & 5.8 (0.8) & 101 (28) \\
 & 241904.643 & 5$_{2,0}$ -- 4$_{2,0}$ & 49.2 & 5.0 & 653 (151) & -10.9 (0.2) & 5.9 (0.7) & 105 (21) \\
 & 242446.084 & 14$_{-1,0}$ -- 13$_{-2,0}$ & 241.0 & 2.3 & 53 (28) & -9.4 (0.9) & 5.4 (2.3) & 9 (3) \\
 & 254015.377 & 2$_{0,0}$ -- 1$_{-1,0}$ & 12.2 & 1.9 & 566 (37) & -12.2 (0.2) & 7.9 (0.4) & 67 (3) \\
 & 261704.409 & 12$_{6,0}$ -- 13$_{5,0}$ & 351.9 & 1.8 & 45 (41) & -10.4 (0.6) & 3.5 (2.9) & 12 (5) \\
 & 261805.675 & 2$_{1,0}$ -- 1$_{0,0}$ & 20.1 & 5.6 & 436 (39) & -11.6 (0.2) & 6.3 (0.4) & 65 (4) \\
 & 265289.562 & 6$_{1,0}$ -- 5$_{2,0}$ & 61.9 & 2.6 & 141 (21) & -11.1 (0.3) & 5.8 (0.7) & 23 (2) \\
 & 266838.148 & 5$_{2,0}$ -- 4$_{1,0}$ & 49.2 & 7.7 & 569 (98) & -11.3 (0.1) & 5.0 (0.4) & 107 (16) \\
 & 267403.471 & 9$_{0,0}$ -- 8$_{1,0}$ & 109.6 & 4.7 & 115 (1,500) & -10.5 (4.6) & 5.1 (22.3) & 21 (263) \\
 & 268743.954 & 9$_{-5,0}$ -- 10$_{-4,0}$ & 220.5 & 1.8 & 35 (19) & -11.7 (0.8) & 3.9 (1.6) & 8 (3) \\
 & 278304.512 & 9$_{-1,0}$ -- 8$_{0,0}$ & 102.1 & 7.7 & 366 (66) & -11.7 (0.2) & 4.7 (0.6) & 73 (9) \\
 & 278599.037 & 14$_{4,0}$ -- 15$_{3,0}$ & 331.8 & 3.6 & 89 (30) & -12.8 (0.8) & 7.2 (1.9) & 12 (2) \\
 & 337135.853 & 3$_{3,0}$ -- 4$_{2,0}$ & 53.7 & 1.6 & 85 (96) & -11.1 (3.2) & 5.2 (5.5) & 16 (5) \\
 & 337642.478 & 7$_{1,1}$ -- 6$_{1,1}$ & 348.4 & 16.5 & 79 (31) & -12.0 (0.6) & 4.8 (1.5) & 15 (4) \\
 & 337643.915 & 7$_{0,1}$ -- 6$_{0,1}$ & 357.5 & 16.9 & 84 (29) & -10.7 (0.6) & 5.0 (1.3) & 16 (4) \\
 & 338124.488 & 7$_{0,0}$ -- 6$_{0,0}$ & 70.2 & 16.9 & 532 (44) & -11.5 (0.1) & 5.0 (0.3) & 100 (5) \\
 & 338344.588 & 7$_{-1,0}$ -- 6$_{-1,0}$ & 62.7 & 16.6 & 769 (68) & -11.4 (0.1) & 3.9 (0.2) & 187 (12) \\
 & 338559.963 & 7$_{-3,0}$ -- 6$_{-3,0}$ & 119.8 & 14.0 & 128 (40) & -10.6 (0.5) & 5.3 (1.3) & 23 (5) \\
 & 338614.936 & 7$_{1,0}$ -- 6$_{1,0}$ & 78.2 & 17.1 & 382 (42) & -11.1 (0.2) & 5.4 (0.4) & 66 (5) \\
 & 338721.693 & 7$_{2,0}$ -- 6$_{2,0}$ & 79.4 & 15.5 & 311 (129) & -11.4 (0.2) & 3.6 (0.8) & 81 (29) \\
 & 338722.898 & 7$_{-2,0}$ -- 6$_{-2,0}$ & 83.0 & 15.7 & 300 (178) & -10.4 (0.3) & 3.6 (1.0) & 78 (41) \\ \hline
\end{tabular}
\label{}
\end{table*}

\begin{table*}[htbp]
\caption{A-CH3OH emission lines from the hot corino detected with the IRAM~30m telescope. Intensities are expressed in units of $T_A^*$.}
\begin{tabular}{rrrrrrrrr}\hline \hline
Species & Frequency & Quantum numbers & $E_{up}$ & $A_{ul}$ & Flux & V$_{lsr}$ & $\Delta$V & $T_{peak}$ \\
 & (MHz) &  & (K) & (10$^{-5}$ s$^{-1})$ & (mK.km.s$^{-1}$) & (km.s$^{-1}$) & (km.s$^{-1}$) & (mK) \\ \hline
A-CH$_3$OH & 95169.391 & 8$_{0,+,0}$ -- 7$_{1,+,0}$ & 83.5 & 0.2 & 618 (22) & -12.4 (0.0) & 4.5 (0.1) & 129 (3) \\
 & 95914.310 & 2$_{1,+,0}$ -- 1$_{1,+,0}$ & 21.4 & 0.2 & 213 (23) & -13.1 (0.3) & 7.0 (0.4) & 28 (3) \\
 & 96741.371 & 2$_{0,+,0}$ -- 1$_{0,+,0}$ & 7.0 & 0.3 & 1,231 (85) & -12.5 (0.2) & 5.2 (0.3) & 223 (9) \\
 & 97582.798 & 2$_{1,-,0}$ -- 1$_{1,-,0}$ & 21.6 & 0.3 & 201 (22) & -12.1 (0.1) & 5.7 (0.4) & 33 (3) \\
 & 132621.824 & 6$_{2,-,0}$ -- 7$_{1,-,0}$ & 86.5 & 0.4 & 46 (8) & -10.3 (0.2) & 4.0 (0.6) & 11 (1) \\
 & 143865.795 & 3$_{1,+,0}$ -- 2$_{1,+,0}$ & 28.3 & 1.1 & 310 (36) & -11.8 (0.1) & 6.0 (0.5) & 48 (4) \\
 & 145103.185 & 3$_{0,+,0}$ -- 2$_{0,+,0}$ & 13.9 & 1.2 & 989 (89) & -12.7 (0.1) & 4.3 (0.2) & 216 (16) \\
 & 146368.328 & 3$_{1,-,0}$ -- 2$_{1,-,0}$ & 28.6 & 1.1 & 494 (25) & -12.0 (0.1) & 5.6 (0.2) & 83 (3) \\
 & 156127.544 & 6$_{2,+,0}$ -- 7$_{1,+,0}$ & 86.5 & 0.7 & 51 (7) & -10.3 (0.1) & 3.5 (0.4) & 14 (1) \\
 & 156602.395 & 2$_{1,+,0}$ -- 3$_{0,+,0}$ & 21.4 & 0.9 & 210 (11) & -10.8 (0.1) & 5.4 (0.2) & 36 (1) \\
 & 201445.493 & 5$_{2,+,0}$ -- 6$_{1,+,0}$ & 72.5 & 1.3 & 132 (52) & -9.5 (0.9) & 6.8 (2.0) & 18 (5) \\
 & 205791.270 & 1$_{1,+,0}$ -- 2$_{0,+,0}$ & 16.8 & 6.3 & 231 (33) & -11.0 (0.3) & 6.9 (0.8) & 31 (3) \\
 & 231281.110 & 10$_{2,-,0}$ -- 9$_{3,-,0}$ & 165.3 & 1.8 & 79 (34) & -10.7 (0.6) & 4.6 (1.5) & 16 (4) \\
 & 232418.521 & 10$_{2,+,0}$ -- 9$_{3,+,0}$ & 165.4 & 1.9 & 81 (34) & -12.2 (0.9) & 6.9 (2.2) & 11 (3) \\
 & 234683.370 & 4$_{2,-,0}$ -- 5$_{1,-,0}$ & 60.9 & 1.8 & 100 (30) & -11.9 (0.6) & 6.3 (1.4) & 15 (3) \\
 & 239746.219 & 5$_{1,+,0}$ -- 4$_{1,+,0}$ & 49.1 & 5.7 & 489 (29) & -11.0 (0.1) & 6.8 (0.3) & 68 (3) \\
 & 241791.352 & 5$_{0,+,0}$ -- 4$_{0,+,0}$ & 34.8 & 6.0 & 2,353 (117) & -11.4 (0.0) & 6.5 (0.2) & 340 (14) \\
 & 241832.718 & 5$_{3,+,0}$ -- 4$_{3,+,0}$ & 84.6 & 3.9 & 296 (40) & -11.0 (0.3) & 6.5 (0.7) & 43 (4) \\
 & 241833.106 & 5$_{3,-,0}$ -- 4$_{3,-,0}$ & 84.6 & 3.9 & 301 (38) & -10.5 (0.2) & 6.5 (0.7) & 43 (3) \\
 & 241842.284 & 5$_{2,-,0}$ -- 4$_{2,-,0}$ & 72.5 & 5.1 & 228 (3) & -11.2 (0.0) & 6.4 (0.1) & 33 (0) \\
 & 243915.788 & 5$_{1,-,0}$ -- 4$_{1,-,0}$ & 49.7 & 6.0 & 336 (119) & -11.0 (0.3) & 5.6 (1.1) & 56 (16) \\
 & 247228.587 & 4$_{2,+,0}$ -- 5$_{1,+,0}$ & 60.9 & 2.2 & 68 (31) & -10.6 (0.8) & 5.4 (1.9) & 12 (3) \\
 & 249887.467 & 14$_{3,-,0}$ -- 14$_{2,+,0}$ & 293.5 & 8.2 & 70 (28) & -9.4 (0.7) & 5.8 (1.7) & 11 (3) \\
 & 250291.181 & 13$_{3,-,0}$ -- 13$_{2,+,0}$ & 261.0 & 8.2 & 57 (20) & -11.9 (0.7) & 5.9 (1.6) & 9 (2) \\
 & 250506.853 & 11$_{0,+,0}$ -- 10$_{1,+,0}$ & 153.1 & 4.2 & 255 (13) & -10.7 (0.1) & 6.6 (0.3) & 36 (1) \\
 & 250635.200 & 12$_{3,-,0}$ -- 12$_{2,+,0}$ & 230.8 & 8.2 & 71 (21) & -11.8 (0.6) & 6.0 (1.4) & 11 (2) \\
 & 250924.398 & 11$_{3,-,0}$ -- 11$_{2,+,0}$ & 203.0 & 8.2 & 65 (25) & -10.7 (0.6) & 4.9 (1.4) & 13 (3) \\
 & 251359.888 & 9$_{3,-,0}$ -- 9$_{2,+,0}$ & 154.3 & 8.1 & 105 (24) & -11.0 (0.4) & 5.8 (1.0) & 17 (3) \\
 & 251517.309 & 8$_{3,-,0}$ -- 8$_{2,+,0}$ & 133.4 & 7.9 & 132 (39) & -10.3 (0.7) & 7.4 (1.9) & 17 (2) \\
 & 251641.787 & 7$_{3,-,0}$ -- 7$_{2,+,0}$ & 114.8 & 7.7 & 115 (19) & -10.3 (0.3) & 5.4 (0.7) & 20 (2) \\
 & 251738.437 & 6$_{3,-,0}$ -- 6$_{2,+,0}$ & 98.5 & 7.4 & 152 (21) & -10.4 (0.2) & 5.2 (0.6) & 27 (2) \\
 & 251984.837 & 8$_{3,+,0}$ -- 8$_{2,-,0}$ & 133.4 & 8.0 & 133 (21) & -10.6 (0.3) & 5.4 (0.7) & 23 (2) \\
 & 252090.409 & 9$_{3,+,0}$ -- 9$_{2,-,0}$ & 154.2 & 8.1 & 116 (21) & -10.2 (0.3) & 5.3 (0.7) & 21 (2) \\
 & 252252.849 & 10$_{3,+,0}$ -- 10$_{2,-,0}$ & 177.5 & 8.3 & 104 (19) & -11.0 (0.3) & 4.8 (0.7) & 20 (3) \\
 & 252485.675 & 11$_{3,+,0}$ -- 11$_{2,-,0}$ & 203.0 & 8.4 & 91 (26) & -10.4 (0.5) & 5.7 (1.2) & 15 (3) \\
 & 253221.376 & 13$_{3,+,0}$ -- 13$_{2,-,0}$ & 261.0 & 8.5 & 66 (22) & -10.7 (0.5) & 4.6 (1.1) & 14 (3) \\
 & 256228.714 & 17$_{3,+,0}$ -- 17$_{2,-,0}$ & 404.8 & 9.0 & 65 (53) & -10.1 (1.4) & 5.6 (4.0) & 11 (4) \\
 & 257402.086 & 18$_{3,+,0}$ -- 18$_{2,-,0}$ & 446.5 & 9.1 & 57 (25) & -11.5 (0.7) & 4.4 (1.5) & 12 (4) \\
 & 279351.887 & 11$_{2,-,0}$ -- 10$_{3,-,0}$ & 190.9 & 3.5 & 42 (22) & -10.0 (0.6) & 4.0 (1.6) & 10 (3) \\
 & 281000.109 & 11$_{2,+,0}$ -- 10$_{3,+,0}$ & 190.9 & 3.5 & 101 (24) & -11.1 (0.3) & 4.4 (0.8) & 22 (3) \\
 & 292672.889 & 6$_{1,-,0}$ -- 5$_{1,-,0}$ & 63.7 & 10.6 & 530 (43) & -10.9 (0.1) & 4.2 (0.3) & 118 (6) \\
 & 302912.979 & 12$_{0,+,0}$ -- 11$_{1,+,0}$ & 180.9 & 7.6 & 310 (36) & -10.6 (0.2) & 4.9 (0.4) & 59 (4) \\
 & 303366.921 & 1$_{1,-,0}$ -- 1$_{0,+,0}$ & 16.9 & 22.6 & 829 (42) & -10.3 (0.1) & 6.5 (0.2) & 120 (4) \\
 & 304208.348 & 2$_{1,-,0}$ -- 2$_{0,+,0}$ & 21.6 & 21.1 & 678 (55) & -10.5 (0.1) & 4.9 (0.3) & 130 (7) \\
 & 305473.491 & 3$_{1,-,0}$ -- 3$_{0,+,0}$ & 28.6 & 16.3 & 645 (97) & -10.5 (0.1) & 4.6 (0.4) & 131 (16) \\
 & 307165.924 & 4$_{1,-,0}$ -- 4$_{0,+,0}$ & 38.0 & 16.5 & 563 (60) & -10.4 (0.3) & 6.0 (0.5) & 88 (5) \\
 & 309290.360 & 5$_{1,-,0}$ -- 5$_{0,+,0}$ & 49.7 & 16.8 & 285 (136) & -10.2 (0.3) & 4.4 (1.0) & 60 (25) \\
 & 331502.319 & 11$_{1,-,0}$ -- 11$_{0,+,0}$ & 169.0 & 19.6 & 230 (48) & -10.6 (0.3) & 4.1 (0.7) & 53 (7) \\
 & 335582.017 & 7$_{1,+,0}$ -- 6$_{1,+,0}$ & 79.0 & 16.3 & 293 (45) & -11.1 (0.2) & 3.8 (0.5) & 72 (7) \\
 & 336865.149 & 12$_{1,-,0}$ -- 12$_{0,+,0}$ & 197.1 & 20.3 & 179 (46) & -11.3 (0.4) & 5.4 (1.1) & 31 (5) \\
 & 337297.484 & 7$_{1,+,1}$ -- 6$_{1,+,1}$ & 390.0 & 16.5 & 80 (81) & -10.3 (1.6) & 4.8 (4.2) & 16 (7) \\
 & 338408.698 & 7$_{0,+,0}$ -- 6$_{0,+,0}$ & 65.0 & 17.0 & 820 (73) & -11.5 (0.1) & 3.8 (0.2) & 200 (13) \\
 & 338486.322 & 7$_{5,+,0}$ -- 6$_{5,+,0}$ & 202.9 & 8.4 & 98 (41) & -12.3 (0.7) & 5.0 (1.6) & 18 (5) \\
 & 338486.322 & 7$_{5,-,0}$ -- 6$_{5,-,0}$ & 202.9 & 8.4 & 104 (43) & -12.4 (0.7) & 5.5 (1.8) & 18 (5) \\
 & 338512.632 & 7$_{4,-,0}$ -- 6$_{4,-,0}$ & 145.3 & 11.5 & 185 (44) & -10.4 (0.3) & 4.7 (0.9) & 37 (5) \\
 & 338512.644 & 7$_{4,+,0}$ -- 6$_{4,+,0}$ & 145.3 & 11.5 & 182 (44) & -10.4 (0.3) & 4.6 (0.9) & 37 (5) \\
 & 338512.853 & 7$_{2,-,0}$ -- 6$_{2,-,0}$ & 102.7 & 15.7 & 180 (44) & -10.2 (0.3) & 4.6 (0.9) & 37 (5) \\
 & 338540.826 & 7$_{3,+,0}$ -- 6$_{3,+,0}$ & 114.8 & 13.9 & 290 (42) & -11.7 (0.2) & 5.4 (0.6) & 50 (5) \\
 & 338543.152 & 7$_{3,-,0}$ -- 6$_{3,-,0}$ & 114.8 & 13.9 & 293 (42) & -9.6 (0.2) & 5.4 (0.6) & 51 (5) \\
 & 338639.802 & 7$_{2,+,0}$ -- 6$_{2,+,0}$ & 102.7 & 15.7 & 124 (40) & -11.0 (0.5) & 4.8 (1.2) & 24 (5) \\
 & 341415.615 & 7$_{1,-,0}$ -- 6$_{1,-,0}$ & 80.1 & 17.1 & 180 (115) & -11.0 (0.3) & 3.3 (1.1) & 52 (28) \\
 & 342729.796 & 13$_{1,-,0}$ -- 13$_{0,+,0}$ & 227.5 & 21.1 & 202 (83) & -10.6 (0.5) & 4.8 (1.8) & 40 (7) \\ \hline
\end{tabular}
\label{}
\end{table*}

\begin{table*}[htbp]
\caption{Rare CH$_3$OH isotopologue emission lines from the hot corino detected with the IRAM~30m telescope. Intensities are expressed in units of $T_A^*$. }
\begin{tabular}{rrrrrrrrr}\hline \hline
Species & Frequency & Quantum numbers & $E_{up}$ & $A_{ul}$ & Flux & V$_{lsr}$ & $\Delta$V & $T_{peak}$ \\
 & (MHz) &  & (K) & (10$^{-5}$ s$^{-1})$ & (mK.km.s$^{-1}$) & (km.s$^{-1}$) & (km.s$^{-1}$) & (mK) \\ \hline
$^{13}$CH$_3$OH & 141602.528 & 3$_{0,3,+}$ -- 2$_{0,2,+}$ & 13.6 & 1.1 & 83 (17) & -11.7 (0.3) & 5.9 (0.8) & 13 (2) \\
 & 168676.499 & 3$_{2,1}$ -- 2$_{1,1}$ & 35.9 & 2.5 & 46 (13) & -11.0 (0.3) & 4.1 (1.0) & 11 (2) \\
 & 235881.170 & 5$_{0,5}$ -- 4$_{0,4}$ & 47.1 & 5.6 & 51 (28) & -9.3 (1.1) & 6.2 (2.6) & 8 (3) \\
 & 235938.220 & 5$_{-1,5}$ -- 4$_{-1,4}$ & 39.6 & 5.4 & 89 (32) & -11.4 (1.0) & 8.8 (2.4) & 10 (2) \\
 & 236049.520 & 5$_{2,3,+}$ -- 4$_{2,2,+}$ & 71.8 & 4.8 & 59 (24) & -10.4 (0.8) & 5.7 (1.8) & 10 (3) \\
 & 255214.891 & 7$_{3,5,+}$ -- 7$_{2,6,-}$ & 113.5 & 8.1 & 38 (23) & -11.7 (0.7) & 4.2 (2.0) & 8 (3) \\
CH$_2$DOH & 88073.074 & 2$_{1,2,0}$ -- 1$_{1,1,0}$ & 10.4 & 0.1 & 13 (4) & -11.3 (0.2) & 2.0 (0.5) & 6 (1) \\
 & 91586.845 & 4$_{1,3,0}$ -- 4$_{0,4,0}$ & 25.8 & 0.5 & 44 (8) & -11.2 (0.2) & 3.7 (0.5) & 11 (1) \\
 & 98031.213 & 4$_{0,4,0}$ -- 3$_{1,3,0}$ & 21.4 & 0.2 & 33 (5) & -11.5 (0.1) & 2.3 (0.3) & 13 (1) \\
 & 132093.628 & 3$_{1,3,0}$ -- 2$_{1,2,0}$ & 16.7 & 0.6 & 43 (6) & -10.9 (0.1) & 2.6 (0.3) & 15 (1) \\
 & 134065.381 & 3$_{0,3,0}$ -- 2$_{0,2,0}$ & 12.9 & 0.7 & 72 (17) & -11.3 (0.3) & 5.5 (1.0) & 12 (2) \\
 & 136151.142 & 3$_{1,2,0}$ -- 2$_{1,1,0}$ & 17.1 & 0.7 & 42 (19) & -10.2 (0.9) & 5.5 (1.7) & 7 (2) \\
 & 161602.516 & 3$_{1,3,1}$ -- 4$_{0,4,0}$ & 29.2 & 0.9 & 52 (19) & -11.0 (0.5) & 5.9 (2.0) & 8 (1) \\
 & 172015.919 & 2$_{1,2,0}$ -- 1$_{0,1,0}$ & 10.4 & 2.2 & 71 (21) & -10.9 (0.3) & 3.8 (0.8) & 17 (4) \\
 & 237249.907 & 7$_{2,5,0}$ -- 7$_{1,6,0}$ & 76.4 & 8.1 & 32 (7) & -12.4 (0.4) & 4.6 (0.8) & 6 (1) \\
 & 240187.202 & 6$_{2,4,0}$ -- 6$_{1,5,0}$ & 61.3 & 7.7 & 163 (17) & -10.3 (0.2) & 6.1 (0.5) & 25 (2) \\
 & 243225.990 & 5$_{2,3,0}$ -- 5$_{1,4,0}$ & 48.4 & 6.7 & 89 (35) & -9.5 (0.7) & 6.0 (1.9) & 14 (3) \\
 & 253637.057 & 4$_{2,2,1}$ -- 3$_{1,2,2}$ & 48.0 & 6.0 & 103 (46) & -10.2 (1.0) & 7.2 (2.5) & 13 (4) \\
 & 255647.816 & 3$_{2,2,0}$ -- 3$_{1,3,0}$ & 29.0 & 6.3 & 154 (44) & -10.9 (0.6) & 6.7 (1.4) & 21 (4) \\
 & 256731.552 & 4$_{1,4,0}$ -- 3$_{0,3,0}$ & 25.2 & 6.9 & 101 (46) & -11.2 (0.6) & 4.4 (1.6) & 22 (6) \\
 & 257895.673 & 4$_{2,3,1}$ -- 3$_{1,3,2}$ & 48.0 & 6.3 & 89 (45) & -10.5 (0.7) & 4.2 (1.6) & 20 (7) \\
 & 264017.721 & 6$_{1,6,0}$ -- 5$_{1,5,0}$ & 48.4 & 5.8 & 131 (31) & -9.9 (0.4) & 5.3 (1.0) & 23 (4) \\
 & 265509.198 & 6$_{1,6,2}$ -- 5$_{1,5,2}$ & 67.5 & 7.5 & 70 (28) & -10.7 (0.6) & 4.2 (1.3) & 16 (4) \\
 & 267031.234 & 6$_{0,6,1}$ -- 5$_{0,5,1}$ & 58.4 & 7.5 & 50 (19) & -10.3 (0.6) & 4.7 (1.3) & 10 (2) \\
 & 267634.613 & 6$_{0,6,0}$ -- 5$_{0,5,0}$ & 45.0 & 6.3 & 155 (33) & -10.9 (0.3) & 4.4 (0.7) & 33 (5) \\
 & 268012.863 & 6$_{2,5,0}$ -- 5$_{2,4,0}$ & 61.2 & 5.4 & 111 (16) & -10.7 (0.3) & 6.9 (0.7) & 15 (1) \\
 & 268059.704 & 6$_{2,4,1}$ -- 5$_{2,3,1}$ & 71.6 & 7.1 & 71 (34) & -11.9 (0.8) & 5.2 (2.0) & 13 (4) \\
 & 269886.317 & 6$_{1,5,2}$ -- 5$_{1,4,2}$ & 68.2 & 7.9 & 50 (23) & -10.2 (0.4) & 3.7 (1.4) & 12 (4) \\
 & 270299.931 & 7$_{2,6,0}$ -- 7$_{1,7,0}$ & 76.2 & 8.8 & 102 (24) & -11.5 (0.4) & 5.7 (1.0) & 17 (3) \\
 & 270734.570 & 6$_{1,5,1}$ -- 5$_{1,4,1}$ & 62.0 & 7.8 & 72 (27) & -11.3 (0.6) & 4.5 (1.3) & 15 (4) \\
 & 272098.060 & 6$_{1,5,0}$ -- 5$_{1,4,0}$ & 49.8 & 6.4 & 96 (35) & -9.9 (0.6) & 5.1 (1.4) & 18 (4) \\
 & 275510.466 & 8$_{2,7,0}$ -- 8$_{1,8,0}$ & 93.3 & 9.4 & 75 (17) & -11.9 (0.2) & 3.7 (0.7) & 19 (3) \\ \hline
\end{tabular}
\label{}
\end{table*}

\begin{table*}[htbp]
\caption{COMs emission lines from the hot corino detected with the IRAM~30m telescope. Intensities are expressed in units of $T_A^*$. }
\begin{tabular}{rrrrrrrrr}\hline \hline
Species & Frequency & Quantum numbers & $E_{up}$ & $A_{ul}$ & Flux & V$_{lsr}$ & $\Delta$V & $T_{peak}$ \\
 & (MHz) &  & (K) & (10$^{-5}$ s$^{-1})$ & (mK.km.s$^{-1}$) & (km.s$^{-1}$) & (km.s$^{-1}$) & (mK) \\ \hline
CH$_3$OCH$_3$ & 132524.779 & 8$_{0,8,0}$ -- 7$_{1,7,0}$ & 32.4 & 1.1 & 46 (12) & -11.6 (0.2) & 2.4 (0.5) & 18 (3) \\
 & 132525.239 & 8$_{0,8,1}$ -- 7$_{1,7,1}$ & 32.4 & 1.1 & 43 (11) & -10.5 (0.2) & 2.1 (0.4) & 19 (3) \\
 & 222247.600 & 4$_{3,2,1}$ -- 3$_{2,1,1}$ & 21.8 & 3.4 & 67 (33) & -11.5 (0.9) & 5.9 (2.2) & 11 (3) \\
 & 222247.600 & 4$_{3,2,3}$ -- 3$_{2,1,3}$ & 21.8 & 4.9 & 68 (28) & -11.7 (0.7) & 5.4 (1.7) & 12 (3) \\
 & 222254.582 & 4$_{3,2,0}$ -- 3$_{2,1,0}$ & 21.8 & 4.9 & 58 (35) & -12.3 (1.0) & 5.5 (2.6) & 10 (4) \\
 & 222426.698 & 4$_{3,1,3}$ -- 3$_{2,2,3}$ & 21.8 & 4.9 & 43 (38) & -9.5 (2.0) & 7.7 (5.7) & 5 (3) \\
 & 223202.244 & 8$_{2,7,1}$ -- 7$_{1,6,1}$ & 38.3 & 3.1 & 39 (37) & -9.1 (0.7) & 4.0 (3.0) & 9 (5) \\
 & 237618.803 & 9$_{2,8,2}$ -- 8$_{1,7,2}$ & 46.5 & 3.8 & 75 (39) & -13.4 (1.1) & 6.4 (2.6) & 11 (4) \\
 & 237618.808 & 9$_{2,8,3}$ -- 8$_{1,7,3}$ & 46.5 & 3.8 & 84 (58) & -13.5 (1.7) & 7.4 (4.4) & 11 (4) \\
 & 237620.887 & 9$_{2,8,1}$ -- 8$_{1,7,1}$ & 46.5 & 3.8 & 81 (37) & -10.9 (1.0) & 7.0 (2.4) & 11 (3) \\
 & 237622.968 & 9$_{2,8,0}$ -- 8$_{1,7,0}$ & 46.5 & 3.8 & 79 (39) & -8.0 (1.0) & 6.5 (2.4) & 11 (4) \\
 & 240989.939 & 5$_{3,3,0}$ -- 4$_{2,2,0}$ & 26.3 & 5.4 & 55 (33) & -11.8 (1.0) & 5.5 (2.4) & 10 (4) \\
 & 258548.775 & 14$_{1,14,2}$ -- 13$_{0,13,2}$ & 93.3 & 10.4 & 177 (53) & -11.3 (0.7) & 7.5 (1.7) & 22 (4) \\
 & 258548.775 & 14$_{1,14,3}$ -- 13$_{0,13,3}$ & 93.3 & 10.4 & 175 (52) & -11.3 (0.7) & 7.3 (1.6) & 23 (4) \\
 & 258549.019 & 14$_{1,14,1}$ -- 13$_{0,13,1}$ & 93.3 & 10.4 & 175 (51) & -10.9 (0.7) & 7.1 (1.6) & 23 (4) \\
 & 258549.263 & 14$_{1,14,0}$ -- 13$_{0,13,0}$ & 93.3 & 10.4 & 180 (53) & -10.7 (0.7) & 7.5 (1.7) & 23 (4) \\
 & 269608.758 & 15$_{0,15,0}$ -- 14$_{1,14,0}$ & 106.3 & 12.0 & 121 (38) & -9.3 (0.8) & 8.7 (2.2) & 13 (2) \\
 & 269608.775 & 15$_{0,15,1}$ -- 14$_{1,14,1}$ & 106.3 & 12.0 & 106 (34) & -9.3 (0.8) & 7.7 (1.9) & 13 (3) \\
 & 269608.791 & 15$_{0,15,3}$ -- 14$_{1,14,3}$ & 106.3 & 12.0 & 109 (37) & -9.4 (0.8) & 7.7 (2.0) & 13 (3) \\
 & 269608.792 & 15$_{0,15,2}$ -- 14$_{1,14,2}$ & 106.3 & 12.0 & 118 (47) & -10.1 (1.1) & 8.5 (2.8) & 13 (3) \\
HCOOCH$_3$ & 89314.657 & 8$_{1,8,1}$ -- 7$_{1,7,1}$ & 20.1 & 1.0 & 18 (12) & -11.4 (0.4) & 2.9 (1.8) & 6 (1) \\
 & 90227.659 & 8$_{0,8,2}$ -- 7$_{0,7,2}$ & 20.1 & 1.1 & 9 (0) & -11.1 (0.0) & 1.6 (0.0) & 5 (0) \\
 & 90229.624 & 8$_{0,8,0}$ -- 7$_{0,7,0}$ & 20.1 & 1.1 & 13 (5) & -11.9 (0.3) & 2.1 (0.6) & 6 (1) \\
 & 103466.572 & 8$_{2,6,2}$ -- 7$_{2,5,2}$ & 24.6 & 1.5 & 17 (12) & -12.0 (0.8) & 2.8 (1.8) & 6 (1) \\
 & 103478.663 & 8$_{2,6,0}$ -- 7$_{2,5,0}$ & 24.6 & 1.5 & 11 (5) & -10.4 (0.2) & 1.8 (0.6) & 5 (2) \\
 & 129296.357 & 10$_{2,8,2}$ -- 9$_{2,7,2}$ & 36.4 & 3.1 & 26 (9) & -11.7 (0.2) & 1.5 (0.4) & 17 (4) \\
 & 132107.205 & 12$_{1,12,0}$ -- 11$_{1,11,0}$ & 42.4 & 3.4 & 18 (5) & -11.1 (0.2) & 1.8 (0.4) & 10 (2) \\
 & 132246.730 & 12$_{0,12,0}$ -- 11$_{0,11,0}$ & 42.4 & 3.4 & 30 (5) & -10.5 (0.1) & 2.8 (0.4) & 10 (1) \\
 & 132921.937 & 11$_{1,10,2}$ -- 10$_{1,9,2}$ & 40.4 & 3.4 & 15 (14) & -11.0 (0.3) & 0.9 (0.7) & 16 (6) \\
 & 132928.736 & 11$_{1,10,0}$ -- 10$_{1,9,0}$ & 40.4 & 3.4 & 22 (10) & -11.7 (0.3) & 1.7 (0.6) & 12 (4) \\
CH$_3$CN & 110364.354 & 6$_{3}$ -- 5$_{3}$ & 82.8 & 8.3 & 41 (19) & -11.4 (0.2) & 3.0 (0.9) & 13 (4) \\
 & 110374.989 & 6$_{2}$ -- 5$_{2}$ & 47.1 & 9.9 & 42 (16) & -10.9 (0.2) & 2.8 (0.7) & 14 (4) \\
 & 128769.436 & 7$_{2}$ -- 6$_{2}$ & 53.3 & 16.4 & 88 (22) & -11.1 (0.2) & 2.8 (0.6) & 29 (4) \\
 & 128779.364 & 7$_{0}$ -- 6$_{0}$ & 24.7 & 17.8 & 293 (30) & -11.3 (0.2) & 5.9 (0.5) & 47 (3) \\
 & 147149.068 & 8$_{3}$ -- 7$_{3}$ & 96.1 & 23.1 & 130 (7) & -11.2 (0.1) & 4.9 (0.2) & 25 (1) \\
 & 147163.244 & 8$_{2}$ -- 7$_{2}$ & 60.4 & 25.2 & 56 (13) & -10.9 (0.1) & 3.1 (0.5) & 17 (3) \\
 & 202320.443 & 11$_{3}$ -- 10$_{3}$ & 122.6 & 65.6 & 115 (41) & -10.5 (0.7) & 5.9 (1.6) & 18 (4) \\
 & 202339.922 & 11$_{2}$ -- 10$_{2}$ & 86.9 & 68.6 & 95 (29) & -10.8 (0.6) & 4.7 (1.1) & 19 (4) \\
 & 202351.612 & 11$_{1}$ -- 10$_{1}$ & 65.4 & 70.3 & 278 (49) & -15.6 (0.5) & 9.4 (1.3) & 28 (3) \\
 & 202355.509 & 11$_{0}$ -- 10$_{0}$ & 58.3 & 70.9 & 280 (48) & -9.7 (0.5) & 9.8 (1.3) & 27 (3) \\
 & 220709.017 & 12$_{3}$ -- 11$_{3}$ & 133.2 & 86.6 & 101 (25) & -10.3 (0.5) & 6.1 (1.2) & 16 (2) \\
 & 220743.011 & 12$_{1}$ -- 11$_{1}$ & 76.0 & 91.8 & 208 (32) & -11.8 (0.5) & 10.1 (1.2) & 19 (2) \\
 & 220747.261 & 12$_{0}$ -- 11$_{0}$ & 68.9 & 92.4 & 234 (33) & -5.5 (0.5) & 11.7 (1.2) & 19 (2) \\
 & 239096.497 & 13$_{3}$ -- 12$_{3}$ & 144.6 & 111.5 & 76 (27) & -10.1 (0.5) & 4.7 (1.3) & 15 (3) \\
 & 239133.313 & 13$_{1}$ -- 12$_{1}$ & 87.5 & 117.2 & 208 (41) & -12.3 (0.6) & 10.1 (1.5) & 19 (2) \\
 & 239137.916 & 13$_{0}$ -- 12$_{0}$ & 80.3 & 117.9 & 213 (41) & -6.6 (0.6) & 10.3 (1.6) & 19 (2) \\
 & 257527.384 & 14$_{0}$ -- 13$_{0}$ & 92.7 & 147.6 & 219 (54) & -9.5 (1.0) & 12.7 (2.4) & 16 (2) \\
 & 275910.263 & 15$_{1}$ -- 14$_{1}$ & 113.1 & 181.1 & 151 (55) & -12.0 (1.1) & 9.4 (2.7) & 15 (3) \\
 & 294279.750 & 16$_{2}$ -- 15$_{2}$ & 148.6 & 217.7 & 74 (28) & -10.5 (0.4) & 3.5 (1.0) & 20 (5) \\
 & 294296.728 & 16$_{1}$ -- 15$_{1}$ & 127.2 & 220.4 & 186 (38) & -11.7 (0.5) & 7.8 (1.2) & 22 (3) \\
\hline
\end{tabular}
\label{}
\end{table*}

\begin{table*}[htbp]
\caption{COMs emission lines from the hot corino detected with the IRAM~30m telescope (continued). Intensities are expressed in units of $T_A^*$.}
\begin{tabular}{rrrrrrrrr}\hline \hline
Species & Frequency & Quantum numbers & $E_{up}$ & $A_{ul}$ & Flux & V$_{lsr}$ & $\Delta$V & $T_{peak}$ \\
 & (MHz) &  & (K) & (10$^{-5}$ s$^{-1})$ & (mK.km.s$^{-1}$) & (km.s$^{-1}$) & (km.s$^{-1}$) & (mK) \\ \hline
H$_2$CCO & 140127.474 & 7$_{1,7}$ -- 6$_{1,6}$ & 40.0 & 3.0 & 85 (21) & -11.1 (0.2) & 5.4 (1.0) & 15 (3) \\
 & 142768.945 & 7$_{1,6}$ -- 6$_{1,5}$ & 40.5 & 3.1 & 112 (15) & -11.3 (0.2) & 6.7 (0.6) & 16 (2) \\
 & 161634.073 & 8$_{0,8}$ -- 7$_{0,7}$ & 34.9 & 4.7 & 86 (45) & -11.4 (0.5) & 6.1 (2.6) & 13 (4) \\
 & 163160.881 & 8$_{1,7}$ -- 7$_{1,6}$ & 48.3 & 4.7 & 133 (23) & -11.5 (0.2) & 5.5 (0.6) & 23 (3) \\
 & 203940.225 & 10$_{1,9}$ -- 9$_{1,8}$ & 66.9 & 9.4 & 69 (20) & -10.6 (0.6) & 4.5 (1.0) & 14 (3) \\
 & 220177.569 & 11$_{1,11}$ -- 10$_{1,10}$ & 76.5 & 11.9 & 41 (35) & -11.7 (1.2) & 3.3 (2.2) & 12 (6) \\
 & 224327.250 & 11$_{1,10}$ -- 10$_{1,9}$ & 77.7 & 12.6 & 69 (21) & -10.8 (0.5) & 4.5 (1.0) & 15 (3) \\
 & 244712.269 & 12$_{1,11}$ -- 11$_{1,10}$ & 89.4 & 16.4 & 50 (34) & -10.6 (0.6) & 3.9 (2.1) & 12 (5) \\
 & 265095.049 & 13$_{1,12}$ -- 12$_{1,11}$ & 102.1 & 21.0 & 50 (27) & -11.1 (0.8) & 5.2 (2.3) & 9 (3) \\
H$_2^{13}$CO & 137449.950 & 2$_{1,2}$ -- 1$_{1,1}$ & 21.7 & 4.9 & 174 (31) & -11.9 (0.2) & 6.3 (0.8) & 26 (3)\\
			 & 141983.740 & 2$_{0,2}$ -- 1$_{0,1}$ & 10.2 & 7.2 & 150 (17) & -11.9 (0.2) & 7.1 (0.5) & 20 (2)\\
			 & 146635.671 & 2$_{1,1}$ -- 1$_{1,0}$ & 22.4 & 6.0 & 173 (31) & -11.8 (0.1) & 5.9 (0.7) & 27 (4)\\
			 & 206131.626 & 3$_{1,3}$ -- 2$_{1,2}$ & 31.6 & 21.1& 892 (96) & -10.8 (0.2) & 6.4 (0.6) & 130 (9)\\
			 & 212811.184 & 3$_{0,3}$ -- 2$_{0,2}$ & 20.4 & 26.1& 201 (43) & -12.4 (0.5) & 7.5 (1.2) & 25 (3)\\
			 & 274762.112 & 4$_{1,4}$ -- 3$_{1,3}$ & 44.8 & 54.7& 259 (56) & -11.1 (0.5) & 7.0 (1.3) & 35 (4)\\
			 & 293126.515 & 4$_{1,3}$ -- 3$_{1,2}$ & 47.0 & 66.4& 222 (46) & -9.8 (0.3)  & 4.9 (0.8) & 42 (6)\\
			 & 343325.713 & 5$_{1,5}$ -- 4$_{1,4}$ & 61.3 &111.8& 290 (85) & -10.8 (0.4) & 4.8 (1.1) & 57 (10)\\  \hline
\end{tabular}
\label{}
\end{table*}
%\end{appendix}

\begin{table}[htbp]
\caption{NOEMA Observations of Cep\,E-A.}
\begin{tabular}{rrrrrrrrr}\hline
Species & Frequency & Quantum numbers & $E_{up}$ & $A_{ul}$ & Flux & V$_{lsr}$ & $\Delta$V & $T_{peak}$ \\
 & (MHz) &  & (K) & (10$^{-5}$ s$^{-1})$ & (mJ/beam.km.s$^{-1}$) & (km.s$^{-1}$) & (km.s$^{-1}$) & (mJ/beam) \\
 \hline\hline
E-CH$_3$OH & 216945.521 & 5$_{1,0}$ -- 4$_{2,0}$ & 48.0 & 1.2 & 807 (37) & -10.3 (0.1) & 5.2 (0.2) & 145.9 (4.5) \\
 & 217886.504 & 20$_{1,0}$ -- 20$_{0,0}$ & 500.5 & 3.4 & 380 (28) & -11.1 (0.1) & 5.5 (0.3) & 65.4 (3.1) \\
 & 218440.063 & 4$_{2,0}$ -- 3$_{1,0}$ & 37.6 & 4.7 & 1330 (49) & -10.4 (0.0) & 4.7 (0.1) & 264.5 (6.1) \\
 & 220078.561 & 8$_{0,0}$ -- 7$_{1,0}$ & 88.7 & 2.5 & 975 (47) & -10.5 (0.1) & 6.3 (0.2) & 146.3 (3.9) \\
A-CH$_3$OH & 217299.205 & 6$_{1,-,1}$ -- 7$_{2,-,1}$ & 373.9 & 4.3 & 504 (29) & -10.8 (0.1) & 5.8 (0.3) & 82.1 (3.1) \\
 & 217642.677 & 15$_{6,-,1}$ -- 16$_{5,-,1}$ & 745.6 & 1.9 & 188 (26) & -12.3 (0.3) & 5.5 (0.6) & 32.4 (3.0) \\
 & 217642.678 & 15$_{6,+,1}$ -- 16$_{5,+,1}$ & 745.6 & 1.9 & 179 (25) & -12.3 (0.2) & 5.3 (0.5) & 31.8 (2.9) \\
CH$_2$DOH & 86668.751 & 2$_{1,1,0}$ -- 2$_{0,2,0}$ & 10.6 & 0.5 & 127 (25) & -11.9 (0.9) & 9.8 (1.4) & 12.1 (1.6) \\
 & 218316.390 & 5$_{2,4,1}$ -- 5$_{1,5,1}$ & 58.7 & 0.9 & 1602 (222) & -12.5 (0.3) & 7.5 (0.9) & 33.8 (2.6) \\
 & 219551.485 & 5$_{1,5,1}$ -- 4$_{1,4,1}$ & 48.2 & 0.7 & 286 (29) & -10.0 (0.3) & 7.9 (0.6) & 33.8 (2.2) \\
 & 220071.805 & 5$_{1,5,0}$ -- 4$_{1,4,0}$ & 35.8 & 3.3 & 1012 (106) & -10.4 (0.2) & 6.6 (0.6) & 56.8 (3.3) \\		
H$_2$CCO & 220177.569 & 11$_{1,11}$ -- 10$_{1,10}$ & 76.5 & 11.9 & 323 (34) & -10.8 (0.2) & 6.6 (0.5) & 46.2 (3.1) \\
HCOOH	& 220038.072 & 10$_{0,10}$ -- 9$_{0,9}$ & 58.6 & 11.5 & 117 (39) & -12.1 (0.8) & 6.0 (1.9) & 13.4 (4.0) \\
HCOOCH$_3$ & 216830.197 & 18$_{2,16,2}$ -- 17$_{2,15,2}$ & 105.7 & 14.8 & 162 (11) & -10.6 (0.2) & 8.4 (0.5) & 18.2 (0.6) \\
 & 216838.889 & 18$_{2,16,0}$ -- 17$_{2,15,0}$ & 105.7 & 14.8 & 114 (8) & -10.2 (0.1) & 4.4 (0.3) & 24.3 (1.1) \\
 & 216964.765 & 20$_{1,20,1}$ -- 19$_{1,19,1}$ & 111.5 & 15.3 & 509 (13) & -12.6 (0.1) & 7.6 (0.2) & 62.8 (1.1) \\
 & 216965.900 & 20$_{1,20,0}$ -- 19$_{1,19,0}$ & 111.5 & 15.3 & 504 (16) & -11.1 (0.1) & 7.5 (0.2) & 62.9 (1.3) \\
 & 216966.246 & 20$_{0,20,2}$ -- 19$_{0,19,2}$ & 111.5 & 15.3 & 500 (49) & -10.6 (0.2) & 7.5 (0.6) & 62.3 (3.9) \\
 & 216967.420 & 20$_{0,20,0}$ -- 19$_{0,19,0}$ & 111.5 & 15.3 & 514 (47) & -9.0 (0.2) & 7.6 (0.5) & 63.2 (3.7) \\
 & 218108.438 & 17$_{4,13,5}$ -- 16$_{4,12,5}$ & 289.7 & 14.8 & 88 (35) & -10.9 (1.0) & 7.9 (2.4) & 10.5 (2.7) \\
 & 218280.900 & 17$_{3,14,2}$ -- 16$_{3,13,2}$ & 99.7 & 15.1 & 147 (29) & -11.5 (0.3) & 5.1 (0.8) & 27.1 (3.5) \\
 & 218297.890 & 17$_{3,14,0}$ -- 16$_{3,13,0}$ & 99.7 & 15.1 & 123 (30) & -11.7 (0.5) & 5.6 (1.0) & 20.8 (3.4) \\
 & 220166.888 & 17$_{4,13,2}$ -- 16$_{4,12,2}$ & 103.2 & 15.2 & 152 (34) & -11.0 (0.5) & 6.5 (1.1) & 21.9 (3.2) \\
 & 220190.285 & 17$_{4,13,0}$ -- 16$_{4,12,0}$ & 103.1 & 15.2 & 140 (49) & -11.9 (0.6) & 6.8 (2.1) & 19.3 (3.3) \\
CH$_3$CHO & 219820.400 & 4$_{2,3,1}$ -- 3$_{1,3,1}$ & 18.3 & 1.5 & 65 (11) & -10.8 (0.2) & 4.5 (0.6) & 13.6 (1.5) \\
CH$_3$OCH$_3$ & 217191.424 & 22$_{4,19,}$ -- 22$_{3,20,}$ & 253.4 & 4.3 & 113 (24) & -11.8 (0.4) & 5.4 (1.1) & 54.0 (4.1) \\
CH$_3$COCH$_3$ & 218127.207 & 20$_{2,18,0}$ -- 19$_{3,17,1}$ & 119.1 & 22.1 & 179 (25) & -11.7 (0.3) & 6.1 (0.6) & 27.4 (2.5) \\
HNCO & 218890 & 10$_{1,10}$ -- 9$_{1,9}$ & 101.1 & 14.6 & 437 (18) & * & 6.7 (0.3) & 56.4 (2.9) \\
	 & 219734 & 10$_{2,9/8}$ -- 9$_{2,8/7}$ & 231.1 & 14.3 & 283 (17) & * & 9.4 (0.8) & 35.3 (2.9) \\
	 & 219798 & 10$_{0,10}$ -- 9$_{0,9}$	& 58.0 & 15.0 & 398 (17) & * & 5.7 (0.5) & 86.3 (2.9) \\
NH$_2$CHO & 218459.653 & 10$_{1,9}$ -- 9$_{1,8}$ & 60.8 & 74.7 & 197 (59) & -12.1 (0.6) & 6.8 (1.7) & 27.1 (4.8) \\
	 \hline

\end{tabular}
 * Unresolved multiplet lines.
\label{}
\end{table}
\end{appendix}

\end{document}